# Satisfiability and Model Checking of CTL* with Graded Path Modalities

**Benjamin Aminof[1], Aniello Murano[2], and Sasha Rubin[1]**

1. TU Vienna, Austria
2. Università degli Studi di Napoli "Federico II", Naples, Italy
   `murano@na.infn.it`

## Abstract

Graded path modalities count the number of paths satisfying a property, and generalize the existential (E) and universal (A) path modalities of CTL*. The resulting logic is called GCTL*. We settle the complexity of satisfiability of GCTL*, i.e., 2ExpTime-Complete, and the complexity of the model checking problem for GCTL*, i.e., PSpace-Complete. The lower bounds already hold for CTL*, and so, using the automata-theoretic approach we supply the upper bounds. The significance of this work is two-fold: GCTL* is more expressive than CTL* at no extra cost in computational complexity, and GCTL* has all the advantages over GCTL (CTL with graded path modalities) that CTL* has over CTL, e.g., the ability to express fairness.

## 1 Introduction

In formal system specification and design, *graded modalities* were introduced as a useful extension of the standard existential and universal quantifiers in branching-time modal logics [5, 6, 11, 16, 21]. These modalities allow one to express properties such as "there exist at least $n$ successors satisfying a formula" or "all but $n$ successors satisfy a formula". The interest in extending logics by graded modalities has been shown to be two-fold: a) typically one gets more expressive logics with very strong forms of counting quantifiers, and b) often the complexity of satisfiability does not increase. For example, this is the case for the extension of $\mu$-calculus by graded (world) modalities, namely $G\mu$-calculus, for which the satisfiability problem remains ExpTime-Complete [16, 6].

Despite its high expressive power, the $\mu$-calculus is a low-level logic, making it "unfriendly" for users. In contrast, more intuitive logics such as CTL and CTL* can naturally express complex properties of computation trees. For this reason, an extension of CTL with graded *path* modalities called GCTL was defined in [4, 5]. Namely, GCTL uses the graded path modalities $\mathsf{E}^{\geq n}\varphi$ and $\mathsf{A}^{<n}\varphi$ that allow one to express properties such as "there are at least $n$ paths satisfying a formula $\varphi$" and "all but at most $n$ paths satisfy a formula $\varphi$", respectively. Although there are several positive results about GCTL this logic suffers from similar limitations as CTL, i.e., it cannot nest successive temporal operators and so cannot express *fairness* constraints. This dramatically limits the usefulness of GCTL, and so we turn instead to GCTL*, an extension of CTL* by graded *path* modalities. Although the syntax and semantics of GCTL* was defined in [5], it has not been explicitly studied, nor are there precise results for the complexity of related decision problems.

The logic GCTL* is useful in several contexts as it allows one to express complex specifications. For example, in a multitasking scheduling design, one can easily express the property "there are at least two computations such that every request is eventually granted" by means of the GCTL* formula $\mathsf{E}^{\geq 2}\mathsf{G}(\textit{request} \rightarrow \textit{request}\ \mathsf{U}\ \textit{granted})$. Clearly this property cannot be expressed in CTL* nor in GCTL. As another example, consider querying XML documents [18]. These documents can be viewed as labeled unranked trees [3] and GCTL* can be used to reason about numbers of links among tags of descendant paths. This is basically due to the fact that CTL$^*$ (but not CTL) subsumes XPATH [12], which is an expressive language for specifying properties of paths in XML-documents, and graded modalities allow one to restrict XPATH queries to a desired number of paths. Actually, we have that GCTL$^*$ is expressively equivalent, over trees, to *monadic path logic* (MPL) [14, 19], which is Monadic Second-Order Logic with set quantification restricted to branches. This reinforces the fact that GCTL$^*$ is an expressive logic.

Before we report our contributions, we discuss, with a small example, one key feature of GCTL$^*$, namely, counting paths. Consider a labeled transition system whose underlying graph is a clique, and suppose that a property of interest is satisfied at the initial node. How should we count the number of paths, starting at the initial node, satisfying the property? Since there are an infinite number of paths starting from the initial node, one possible definition is that this count is infinity. On the other hand, since the property is already satisfied in the initial node, another possible definition is that this count is 1. The definition of GCTL$^*$ from [5], and the one that we use, considers the number of "minimal" paths satisfying the property, and hence in the example the count is 1. Although this definition is similar to that used in GCTL, reasoning about GCTL$^*$ is much more complicated. Indeed, in GCTL temporal operators are always coupled with branching modalities and so the counting involves one temporal operator at the time. In GCTL$^*$, instead, one needs to count paths satisfying nested temporal operators, and this is reflected in the complexity of the machinery we use to solve the satisfiability and model checking problems.

**Our contributions.** In this paper we address the complexity of the satisfiability and model checking problems for GCTL$^*$. Our main result is that satisfiability is 2ExpTime-Complete, and we extract from it the secondary result that model checking is PSpace-Complete. Thus, in both cases, the problem for GCTL$^*$ is not harder than the the one for CTL$^*$. Our upper bounds are obtained by exploiting an automata-theoretic approach for branching-time logics, following [17]; this approach is suitable because GCTL$^*$ turns out to have the tree model property. In broad outline our approach consists of two steps. In the first step we introduce a suitable variant of hesitant alternating tree automata (HTA) called *Graded Hesitant Tree Automata* (GHTA), and reduce the satisfiability problem of GCTL$^*$ to the non-emptiness problem of an exponentially larger GHTA. Unlike HTA, our GHTA can work on finitely-branching trees, not just $k$-ary branching trees. The non-trivial part of this reduction is dealing with the graded path modalities $\mathsf{E}^{\geq g}\psi$, which is harder than dealing with graded world modalities (as in $G\mu$-calculus) and harder than the case in which temporal operators in $\psi$ are always coupled with path quantifiers (as in GCTL). In the second step we show that GCTL$^*$ has the bounded-tree model property (in particular, a satisfiable formula is satisfied in a tree whose branching degree is at most exponential in the size of the formula), and thus we reduce emptiness of GHTA to emptiness of HTA. All missing proofs appear in the Appendix.

**Related Work.** Counting modalities were first introduced by Fine [11] under the name *graded (world) modalities*. A systematic treatment of the complexity of various graded modal logics followed, see for instance [2, 15]. In [16, 6] graded world modalities over $\mu$-calculus was investigated. The work closest to ours concerns graded path modalities over CTL[4, 5]. Graded path modalities over CTL were also studied in [10], but under a different semantics than GCTL. There, the authors also consider some sort of "minimal" paths. However, the semantics given in [10] is tailored for extending only CTL, and it is unclear if and how one can extend their work to CTL$^*$. Expressiveness of graded world modalities over CTL$^*$, i.e. the extension of CTL$^*$ by the ability to say "there exist at least $n$ successors satisfying a certain formula", was studied in [19] using the composition method. It is unclear if that method can yield the complexity bounds we achieve.

## 2 The GCTL* temporal logic

In this section we define the syntax and semantics of the main object of study, GCTL*, an extension of the classical branching-time temporal logic CTL* by *(graded) path quantifiers* of the form $\mathsf{E}^{\geq g}$. We follow the definition of GCTL* from [5, Section 3], but give a slightly simpler syntax. We assume the reader is familiar with the syntax and semantics of the classic temporal logics CTL*, LTL, and CTL. We first introduce some notation.

Models of GCTL* are *Labelled Transition Systems (LTS)* $\mathsf{S} = \langle \Sigma, S, E, \lambda \rangle$ where $\Sigma$ is a set of *labels*, $S$ is a countable set of *states*, $E \subseteq S \times S$ is the *transition relation*, and $\lambda : S \mapsto \Sigma$ is the *labeling function*. Typically $\Sigma = 2^{\mathrm{AP}}$ where AP is a set of *atoms* and every $\lambda(s)$ is a finite subset of AP. For simplicity of presentation we assume that $E$ is total, i.e., for every $q \in S$ there exists $q' \in S$ such that $(q, q') \in E$. The set of finite and infinite paths in $\mathsf{S}$ that start with $q$ is written $\mathsf{pth}(\mathsf{S}, q)$. Define the prefix ordering on paths $\preceq$ on sets of the form $\mathsf{pth}(\mathsf{S}, q)$. If $X \subseteq \mathsf{pth}(\mathsf{S}, q)$ is a set of paths with the same starting element, define $\mathsf{min}(X)$ as the set of minimal elements in $X$ with respect to the prefix ordering $\preceq$.

A $\Sigma$-*labeled tree* $\mathsf{T}$ is a pair $\langle T, V \rangle$ where $T \subseteq \mathbb{N}^*$ is a $\prec$-downward closed set of strings over $\mathbb{N}$ (i.e., here $\prec$ is the prefix ordering on $\mathbb{N}^*$), and $V : T \to \Sigma$ is a labeling, where $\Sigma$ is a set of *labels*. In particular the empty string $\epsilon$ is the root of the tree. For $t \in T$ write $sons(t) \subset T$ for the set $\{s \in T : t \prec s \wedge \neg \exists z. t \prec z \prec s\}$. If every node $v$ of tree $T$ has finite degree, then we say that $T$ is *finitely branching*. If there is a $k \in \mathbb{N}$ such that every node of $v$ of $T$ has degree at most $k$, then say that $T$ is *boundedly branching* or *has branching degree k*. We implicitly view a tree $\mathsf{T} = \langle T, V \rangle$ as the LTS $\langle \Sigma, T, E, V \rangle$ where $(t, s) \in E$ iff $s \in sons(t)$.

▶ **Definition 1** (Syntax of GCTL*). Fix a set of atoms AP. *The GCTL* *state ($\varphi$) and path ($\psi$) formulas are built inductively from AP using the following grammar:*

$$\varphi ::= p \mid \neg\varphi \mid \varphi \vee \varphi \mid \mathsf{E}^{\geq g}\psi \qquad\qquad \psi ::= \varphi \mid \neg\psi \mid \psi \vee \psi \mid \mathsf{X}\psi \mid \psi\mathsf{U}\psi \mid \psi\mathsf{R}\psi$$

In the first item, $p$ varies over AP and $g$ varies over $\mathbb{N}$ (and thus, technically, there are infinitely many rules in this grammar). As usual, $\mathsf{X}, \mathsf{U}$ and $\mathsf{R}$ are called *temporal operators* and $\mathsf{E}^{\geq g}$ (for $g \in \mathbb{N}$) are called *path quantifiers*. The class of GCTL* *formulas* is the set of state formulas generated by the above grammar. The simpler class of *Graded* CTL *formulas* (GCTL) is obtained by requiring each temporal operator occurring in a formula to be immediately preceded by a path quantifier, as in the classical definition of CTL. The class of LTL formulas is obtained as the path formulas in which no path quantifier appears.

The *degree* of the quantifier $\mathsf{E}^{\geq g}$ is the number $g$. For a state formula $\varphi$, we define the *degree* $\mathsf{deg}(\varphi)$ of $\varphi$ as the maximum natural number $g$ occurring among the degrees of all its path quantifiers. The *length* of a formula $\varphi$, denoted by $|\varphi|$, is defined inductively on the structure of $\varphi$ as usual, and using $|\mathsf{E}^{\geq g}\psi|$ equal to $g + 1 + |\psi|$. Clearly, $\mathsf{deg}(\varphi) = \mathcal{O}(|\varphi|)$.

The semantics of GCTL* are defined for labelled transition systems $\mathsf{S}$. The GCTL* formula $\mathsf{E}^{\geq g}\psi$, for GCTL* path formula $\psi$, can be read as *"there exist at least g (minimal $\psi$-conservative) paths"*. Minimality was defined above, and so we now say, informally, what it means for a path to be $\psi$-conservative. An infinite path of $\mathsf{S}$ is $\psi$-conservative if it satisfies $\psi$, and a finite path of $\mathsf{S}$ is $\psi$-conservative if all its (finite and infinite) extensions in $\mathsf{S}$ satisfy $\psi$. Note that this notion uses a semantics of GCTL* over finite paths, and thus the semantics of GCTL* needs to be defined for finite paths (as well as infinite paths). As in [5, Section 3.2] we use the weak-version of semantics of temporal operators for finite paths (defined in [8, Page 30]). Intuitively, temporal operators are interpreted pessimistically (with respect to possible extensions of the path), e.g., $(S, \pi) \models X\psi$ iff $|\pi| \geq 2$ and $(S, \pi_{\geq 1}) \models \psi$. We now define the semantics of GCTL$^*$. Precisely, we just give it for $E^{\geq g}\psi$ as the semantics for the remaining GCTL$^*$ state and path formulas is as usual in CTL$^*$ (see Appendix A).

▶ **Definition 2** (Semantics of $E^{\geq g}$, and $\psi$-conservative). Fix an LTS S and $s \in S$. Then $(S, s) \models E^{\geq g}\psi$, for $g$ a natural number and $\psi$ a GCTL$^*$ path formula, iff the cardinality of the set $\min(Con(S, s, \psi))$ is at least $g$, where $Con(S, s, \psi) := \{\pi \in \mathsf{pth}(S, s) \mid \forall \pi' \in \mathsf{pth}(S, s) : \pi \preceq \pi' \text{ implies } (S, \pi') \models \psi\}$. The paths in $Con(S, s, \psi)$ are called $\psi$-*conservative (in* S *starting at s)*, and paths in $\min(Con(S, s, \psi))$ are called *minimal $\psi$-conservative.*

▶ **Remark.** **1.** The additional operators present in the syntax of GCTL$^*$ in [5], namely $\wedge, \tilde{X}, \tilde{R}, \tilde{U}$ and $A^{<g}$ are dual to the operators $\vee, X, U, R$ and $E^{\geq g}$ respectively. This point is a little subtle for temporal operators, but is easy to see once one understands that the tilded operators are defined like the untilded operators on infinite paths, but with 'optimistic semantics' on finite paths; e.g., $(S, \pi) \models \tilde{X}\psi$ :if $|\pi| = 1$ or $(S, \pi_{\geq 1}) \models \psi$.
**2.** CTL$^*$ is a fragment of GCTL$^*$ in the following sense: a CTL$^*$ formula $\varphi$ is equivalent to the GCTL$^*$ formula $\varphi'$ in which every occurence of E is replaced by $E^{\geq 1}$. Similarly, LTL is a fragment of GCTL$^*$ as any LTL formula is a GCTL$^*$ path formula.
**3.** For a state formula $\phi$, the GCTL$^*$ formula $E^{\geq n}X\phi$ expresses that there exist at least $n$ immediate successors of the current node satisfying $\phi$. This uses the following facts: i) a path of length 1 does not satisfy the path formula $X\phi$, and thus is not $X\phi$-conservative; ii) if $(S, \pi) \models X\phi$, then already the prefix of $\pi$ of length 2 satisfies $X\phi$, and thus $\pi$ is minimal $X\phi$-conservative iff $|\pi| = 2$. Hence, $(S, s) \models E^{\geq n}X\phi$ iff there are at least $n$ minimal $X\phi$-conservative paths, which by the facts must all be of length 2, iff there are at least $n$ immediate successors of $s$ satisfying $\phi$.
**4.** GCTL$^*$ is not invariant under bisimulation. Indeed, the formula $E^{\geq 2}Xp$ is false in the tree whose root has exactly one successor satisfying $p$, but true in the bisimilar tree with exactly two successors satisfying $p$.

We now explain how to think of a GCTL$^*$ path formula $\psi$ over atoms AP as an LTL formula $\Psi$ over atoms which themselves are GCTL$^*$ state formulas. This idea, in the context of CTL$^*$, can be found in [17, Proof of Theorem 5.3], and will be used in our subsequent proofs. A formula $\varphi$ is a *state sub-formula* of $\psi$ if i) $\varphi$ is a state formula, and ii) $\varphi$ is a sub-formula of $\psi$. A formula $\varphi$ is a *maximal state sub-formula* of $\psi$ if $\varphi$ is a state sub-formula of $\psi$, and $\varphi$ is not a proper sub-formula of any other state sub-formula of $\psi$. Let $max(\psi) = \{\varphi \mid \varphi \text{ is a maximal state sub-formula of } \psi\}$, and let $\overline{max(\psi)} = \bigcup_{\varphi \in max(\psi)} \{\varphi, \neg\varphi\}$ be the set of all maximal state sub-formulas of $\psi$ and their negations. Every GCTL$^*$ path formula $\psi$ can be viewed as the formula $\Psi$ whose atoms are elements of $max(\psi)$. Note that $\Psi$ is an LTL formula. For example, for $\psi = ((Xp) \cup (E^{\geq 2}Xq)) \vee p$, the state sub-formulas are $\{p, q, E^{\geq 2}Xq\}$, and $max(\psi) = \{p, E^{\geq 2}Xq\}$, and thus $\Psi$ is the LTL formula $(X\underline{p} \cup \underline{E^{\geq 2}Xq}) \vee \underline{p}$ over the atoms $\{p, E^{\geq 2}Xq\}$ (here we underline sub-formulas that are treated as atoms). Given an LTS $S = \langle 2^{\mathrm{AP}}, S, E, \lambda \rangle$ and a GCTL$^*$ path formula $\psi$, we define the *relabeling* of the LTS S by the values of the formulas in $max(\psi)$ as $S_\psi = \langle max(\psi), S, E, L \rangle$ where $L(s)$ is the union of $\lambda(s)$ and the set of $\varphi \in max(\psi)$ such that $(S, s) \models \varphi$. An easy induction shows:

▶ **Lemma 3.** *For every* GCTL$^*$ *path formula $\psi$ over atoms* AP *there is an* LTL *formula $\Psi$ over atoms $max(\psi)$ such that for all S and all paths $\pi$ in S:* $(S, \pi) \models \psi$ *iff* $(S_\psi, \pi) \models \Psi$.

We conclude this section by reporting some useful properties of GCTL$^*$.

▶ **Theorem 4.** **1.** GCTL$^*$ *is invariant under unwinding.*

**2.** GCTL$^*$ *is equivalent, over trees, to Monadic Path Logic. In particular,* GCTL$^*$ *is more expressive than* CTL$^*$.
**3.** GCTL$^*$ *has the finitely-branching tree model property, i.e., if a* GCTL$^*$ *formula $\varphi$ is satisfiable then it is satisfiable in a finitely-branching tree.*[1].

**Proof.** The proof of 1 is standard: to treat $\mathsf{E}^{\geq g}$ instead of $\mathsf{E}$ use the standard $\prec$-preserving bijection between paths in an LTS and paths in its unwinding, and note that the semantics of $\mathsf{E}^{\geq g}$ involve reasoning about $\prec$. The proof of 2 follows from the fact that CTL$^*$ extended with $\mathsf{E}^{\geq g}\mathsf{X}$ operators is equivalent to MPL [19], and that GCTL$^*$ can be expressed in MPL. The proof of 3 is folklore for MPL.[2] ◀

## 3 Graded Hesitant Tree Automata

In this section we discuss the different kinds of automata that we use in this work. First, we make use of classical nondeterministic Büchi word automata (NBW), and nondeterministic finite word automata (NFW). We write $\langle \Sigma, Q, q_0, \delta, G \rangle$ for NBWs and $\langle \Sigma, Q, q_0, \delta, F \rangle$ for NFWs where $\Sigma$ is the input alphabet, $Q$ is the set of states, $q_0$ is the initial state, $\delta \subseteq Q \times \Sigma \times Q$ is the transition relation, $G \subseteq Q$ is the set of accepting states and $F \subseteq Q$ the set of final states.

Alternating tree automata are a generalization of nondeterministic tree automata. Intuitively, a nondeterministic tree automaton visiting a node of the input tree sends at most one copy of itself to each of the sons of the node, whereas an alternating automaton can send several copies of itself to the same son. The automata-theoretic approach to branching time temporal logics uses alternating hesitant tree automata (AHTA), and the reader is referred to [17, Section 5.1], or Appendix B, for detailed definitions. Graded hesitant tree automata (GHTA), which we define below, are a generalisation of AHTA. Here we simply introduce enough notation for AHTA in order to define GHTA.

For a set $X$, let $\mathsf{B}^+(X)$ be the set of positive Boolean formulas over $X$, including the constants **true** and **false**. A set $Y \subseteq X$ satisfies a formula $\theta \in \mathsf{B}^+(X)$, written $Y \models \theta$, if assigning **true** to elements in $Y$ and **false** to elements in $X \setminus Y$ makes $\theta$ true.

An *Alternating Hesitant Tree Automaton* (AHTA) is a tuple $\mathsf{A} = \langle \Sigma, D, Q, q_0, \delta, \langle G, B \rangle,$ $\langle \mathsf{part}, \mathsf{type}, \preceq \rangle \rangle$ where $\Sigma$ is a non-empty finite set of *input letters*; $D \subset \mathbb{N}$ is a finite non-empty set of *directions*, $Q$ is the non-empty finite set of *states*, $q_0 \in Q$ is the *initial state*; the pair $\langle G, B \rangle \in 2^Q \times 2^Q$ is the *acceptance condition* (we sometimes call the states in $G$ *good states* and the states in $B$ *bad states*); $\delta : Q \times \Sigma \mapsto \mathsf{B}^+(D \times Q)$ is the *alternating transition function*; $\mathsf{part} \subset 2^Q$ is a partition of $Q$, $\mathsf{type} : \mathsf{part} \to \{trans, exist, univ\}$ is a function assigning the label *transient*, *existential* or *universal* to each element of the partition, and $\preceq \subset 2^Q \times 2^Q$ is a partial order on $\mathsf{part}$. Moreover, the transition function $\delta$ is required to satisfy the following *hesitancy condition*: for every $\mathbb{Q} \in \mathsf{part}$, every $q \in \mathbb{Q}$, and every $\sigma \in \Sigma$:
(i) for every $\mathbb{Q}' \in \mathsf{part}$ and $q' \in \mathbb{Q}'$, if $q'$ occurs in $\delta(q, \sigma)$ then $\mathbb{Q}' \preceq \mathbb{Q}$,
(ii) if $\mathsf{type}(\mathbb{Q}) \in trans$ then no state of $\mathbb{Q}$ occurs in the formula $\delta(q, \sigma)$,
(iii) if $\mathsf{type}(\mathbb{Q}) \in exist$ (resp., $\mathsf{type}(\mathbb{Q}) \in univ$) then there is at most one element of $\mathbb{Q}$ in each disjunct of the DNF (resp., conjunct of CNF) of $\delta(q, \sigma)$.

An *input tree (for AHTA)* is a $\Sigma$-labeled tree $\mathsf{T} = \langle T, V \rangle$ with $T \subseteq D^*$. Since $D$ is finite, such trees have fixed finite branching degree. A *run* of an alternating tree automaton $\mathsf{A}$ on input tree $\mathsf{T} = \langle T, V \rangle$ is a $(T \times Q)$-labeled tree $\langle T_r, r \rangle$, such that *(i)* $r(\varepsilon) = (\varepsilon, q_0)$ and *(ii)* for all $y \in T_r$, with $r(y) = (x, q)$, there exists a *minimal* set $\mathsf{S} \subseteq D \times Q$, such that $\mathsf{S} \models \delta(q, V(x))$, and for every $(d, q') \in \mathsf{S}$, it is the case that $x \cdot d$ is a son of $x$, and there exists a son $y'$ of $y$, so that $r(y') = (x \cdot d, q')$.

---

[1] Note that we do not state here that there is a bound on the number of children of every node. However, we later prove that such a bound exists, and is exponential in the size of the formula.
[2] We thank Igor Walukiewicz for pointing out to us that this immediately follows from [25, Lemma 43].

The hesitancy condition ensures that every infinite sequence of states of the automaton "gets trapped" in some $\mathbb{Q}$ with $\mathsf{type}(\mathbb{Q}) \in \{exist, univ\}$. We say that an infinite path *gets trapped in an existential set* if $\mathsf{type}(\mathbb{Q}) = exist$, and otherwise we say that it *gets trapped in a universal set*. An infinite path that gets trapped in an existential (resp. universal) set is *accepting* (with respect to $\langle G, B \rangle$) iff it visits some state of $G$ infinitely often (resp. no states from $B$ infinitely often). A run tree is *accepting* if all its infinite paths are accepting.

The *membership problem* of AHTA is the following decision problem: given an AHTA $\mathsf{A}$ with direction set $D$, and a finite LTS $\mathsf{S}$ in which the degree of each node is at most $|D|$, decide whether or not $\mathsf{A}$ accepts $\mathsf{S}$. The *depth* of the AHTA is the size of the longest chain in the partial order $\prec$. The *size* $||\delta||$ of the transition function is the sum of the lengths of the formulas it contains. The *size* $||\mathsf{A}||$ of the AHTA $\mathsf{A}$ is $|D| + |Q| + ||\delta||$. Note that the partition, partial order and type function are not counted in the size of the automaton. The following is implicit in [17, Theorem 5.6].

▶ **Theorem 5.** *The membership problem for AHTA can be solved in space $O(\partial \log^2(|S| \times ||\mathsf{A}||))$ where $\partial$ is the depth of $\mathsf{A}$, $||\mathsf{A}||$ is the size of $\mathsf{A}$, and $S$ is the state set of $\mathsf{S}$.*

We now introduce *Graded Hesitant Tree Automata* (GHTA). These can run on finitely-branching trees (not just trees of a fixed finite degree), and the transition function is graded, i.e., instead of a Boolean combination of direction-state pairs, it specifies a Boolean combination of distribution operations. There are two distribution operations: $\Diamond(q_1,...,q_k)$ and its dual $\Box(q_1,...,q_k)$. Intuitively, $\Diamond(q_1,...,q_k)$ specifies that the automaton picks $k$ *different* sons $s_1,...,s_k$ of the current node and, for each $i \leq k$, sends a copy in state $q_i$ to son $s_i$.

Formally, a GHTA is a tuple $A = \langle \Sigma, Q, q_0, \delta, \langle G, B \rangle, \langle \mathsf{part}, \mathsf{type}, \preceq \rangle \rangle$ where all elements but $\delta$ are defined as for AHTA, and $\delta : Q \times \Sigma \mapsto \mathsf{B}^+(\Diamond_Q \cup \Box_Q)$ is a transition function that maps a state and an input letter to a positive Boolean combination of elements in $\Diamond_Q = \{\Diamond(q_1,...,q_k) \mid (q_1,...,q_k) \in Q^k, k \in \mathbb{N}\}$ and $\Box_Q = \{\Box(q_1,...,q_k) \mid (q_1,...,q_k) \in Q^k, k \in \mathbb{N}\}$.

We now show how to define the run of a GHTA $\mathsf{A}$ on a $\Sigma$-labeled finitely-branching tree $\mathsf{T} = \langle T, V \rangle$ by (locally) unfolding every $\Diamond_Q$ and $\Box_Q$ in $\delta(q, V(t))$ into a formula of $\mathsf{B}^+([d] \times Q)$ where $d$ is the branching-degree of node $t$. For $k, d \in \mathbb{N}$, let $S(k, d)$ be the set of all ordered different $k$ elements in $[d]$, i.e., $(s_1,...,s_k) \in S(k, d)$ iff for every $i \in [k]$ we have that $s_i \in [d]$, and that if $i \neq j$ then $s_i \neq s_j$. Observe that if $k > d$ then $S(k, d) = \emptyset$. For every $d \in \mathbb{N}$, define the function $expand_d : \mathsf{B}^+(\Diamond_Q \cup \Box_Q) \to \mathsf{B}^+([d] \times Q)$ that maps formula $\phi$ to the formula formed from $\phi$ by replacing every occurrence of a sub-formula of the form $\Diamond(q_1,...,q_k)$ by the formula $\bigvee_{(s_1,...,s_k) \in S(k,d)}(\bigwedge(s_i, q_i))$, and every occurrence of a sub-formula of the form $\Box(q_1,...,q_k)$ by the formula $\bigwedge_{(s_1,...,s_k) \in S(k,d)}(\bigvee(s_i, q_i))$. Observe that if $k > d$ then $\Diamond(q_1,...,q_k)$ becomes the constant formula **false**, and $\Box(q_1,...,q_k)$ becomes the constant formula **true**.

The *run of a GHTA* $\mathsf{A}$ is defined as for an alternating tree automaton, except that one uses $expand_n(\delta(q, V(x)))$ instead of $\delta(q, V(x))$ for nodes $x$ of $\mathsf{T}$ of degree $n$.

Finally, the *hesitancy condition* defined above for AHTA are required to apply to the expanded transition function, i.e., insert the phrase "every $n \in \mathbb{N}$," before the phrase "and every $\sigma \in \Sigma$", and in items (i)-(iii) replace $\delta(q, \sigma)$ by $expand_n(\delta(q, \sigma))$. Acceptance is defined as for AHTA.

▶ **Lemma 6.** *The emptiness problem for GHTA* A *over trees of branching degree at most d is decidable in time* $2^{O(d \times |Q|^3)}$, *where Q is the state set of* A.

**Proof.** Convert GHTA A with state set $Q$ into an AHTA with state set $Q$, and note that AHTA are a special case of APTA with 3 priorities. Now apply the fact that emptiness problem for APTA with $p$ priorities over $d$-ary trees can be solved in time $2^{O(d \times |Q|^p)}$ [9]. ◀

## 4 From GCTL* to Graded Hesitant Automata

A standard way for solving the satisfiability and model-checking problems of CTL* is to use the automata theoretic approach [17]. Using this approach, one reduces satisfiability to the non-emptiness problem of a suitable tree automaton accepting all tree models of a given temporal logic formula. We follow the same approach here, by reducing the satisfiability problem of GCTL* to the non-emptiness problem of GHTA. By Theorem 4, a GCTL* formula is satisfiable iff it has a finitely branching (though possibly unboundedly branching) tree model, which exactly falls within the abilities of GHTA. Our main technical result states that every GCTL* formula can be compiled into an exponentially larger GHTA (the rest of this section provides a proof of this result):

▶ **Theorem 7.** *Given a* GCTL* *formula* $\vartheta$, *one can build a GHTA* $A_\vartheta$ *that accepts all the finitely-branching tree models of* $\vartheta$. *Moreover,* $A_\vartheta$ *has* $2^{O(|\vartheta| \times \deg(\vartheta))}$ *states, depth* $O(|\vartheta|)$, *and transition function of size* $2^{O(|\vartheta| \times \deg(\vartheta))}$.

An important observation that allows us to achieve an optimal construction is the following. Suppose that the formula $\mathsf{E}^{\geq g}\psi$ holds at some node $w$ of a tree. Then, by definition, there are at least $g$ different paths $\rho'^1,...,\rho'^g \in \min(Con(\mathsf{S}, w, \psi))$. Look at any $g$ infinite extensions $\rho^1,...,\rho^g$ of these paths in the tree, and note that by the definition of $\psi$-conservativeness all these extensions must satisfy $\psi$. Also observe that for every $i \neq j$, the fact that $\rho'^i, \rho'^j$ are different and minimal implies that the longest common prefix $\rho'^{ij}$ of $\rho^i$ and $\rho^j$ is not $\psi$-conservative. As it turns out, the other direction is also true, i.e., if there are $g$ infinite paths $\rho^1,...,\rho^g$ satisfying $\psi$, such that for every $i \neq j$ the common prefix $\rho'^{ij}$ is not $\psi$-conservative, then there are $g$ prefixes $\rho'^1,...,\rho'^g$ of $\rho^1,...,\rho^g$ respectively, such that $\rho'^1,...,\rho'^g \in \min(Con(\mathsf{S}, w, \psi))$. Note that this allows us to reason about the cardinality of the set $\min(Con(\mathsf{S}, w, \psi))$, by considering only the infinite paths $\rho^1,...,\rho^g$ and their common prefixes, without actually looking at the minimal $\psi$-conservative paths $\rho'^1,...,\rho'^g$. In reality, we do not even have to directly consider the common prefixes $\rho'^{ij}$. Indeed, since the property of being $\psi$-conservative is upward closed (with respect to the prefix ordering $\preceq$ of paths), showing that $\rho'^{ij}$ is not $\psi$-conservative can be done by finding any extension of $\rho'^{ij}$ that is not $\psi$-conservative. The following Proposition formally captures the above.

▶ **Proposition 8.** *Given a* GCTL* *path formula* $\psi$ *and a* $2^{\mathrm{AP}}$-*labeled tree* $\mathsf{T} = (T, V)$, *then* $\mathsf{T} \models \mathsf{E}^{\geq g}\psi$ *iff there are* $g$ *distinct nodes* $y_1,...,y_g \in \mathsf{T}$ (*called* breakpoints) *such that for every* $1 \leq i, j \leq g$ *we have: (i) if* $i \neq j$ *then* $y_i$ *is not a descendant of* $y_j$; *(ii) the path from the root to the father* $x_i$ *of* $y_i$ *is not* $\psi$-*conservative; (iii) there is an infinite path* $\rho^i$ *in* $\mathsf{T}$, *starting at the root and going through* $y_i$, *such that* $\rho^i \models \psi$.

We are now in a position to describe our construction of a GHTA accepting all finitely branching tree models of a given GCTL* formula. The construction, which is detailed below, follows in the footsteps of the one given in [17] for building an AHTA for a CTL* formula. Naturally, the main difference and difficulty lies in handling the graded modalities. The basic intuition behind the way our construction handles formulas of the form $\varphi = \mathsf{E}^{\geq g}\psi$ is the following. Given an input tree, the automaton $\mathsf{A}_\varphi$ for this formula has to find at least $g$ minimal $\psi$-conservative paths. At its core, $\mathsf{A}_\varphi$ runs $g$ pairs of copies of itself in parallel. The reason that these copies are not run independently is to ensure that the two members of each pair are kept coordinated, and that different pairs do not end up making the same guesses (and thus over-counting the number of minimal $\psi$-conservative paths). The task of each of the $g$ pairs is to detect some minimal $\psi$-conservative path that contributes 1 to the count towards $g$. This is done indirectly by using the characterization given by Proposition 8. Since this proposition requires checking whether or not certain paths satisfy $\psi$, the automaton $\mathsf{A}_\varphi$ will access certain classic NBWs. The following theorem follows from [24, 22], and the subsequent lemma is an easy consequence:

▶ **Theorem 9.** *Given an* LTL *formula* $\Psi$, *there are an NBW* $\mathbb{A}_\Psi$ *and an NFW* $\mathbb{B}_\Psi$ *(both of size $2^{O(\Psi)}$) accepting exactly all infinite and finite words that satisfy* $\Psi$, *respectively.*

▶ **Lemma 10.** *Given an* LTL *formula* $\Psi$, *there is an NBW* $\mathbb{A}^\Psi$ *(of size $2^{O(\psi)}$) such that* $\mathbb{A}^\Psi$ *accepts a word* $w$ *iff* $w \models \Psi$, *or* $u \models \Psi$ *for a prefix* $u$ *of* $w$. *Moreover,* $\mathbb{A}^\Psi$ *has an accepting sink* $\top$, *such that if* $r_0, r_1, ...$ *is an accepting run of* $\mathbb{A}^\Psi$ *on* $w$, *and* $i \geq 0$ *satisfies* $r_i \neq \top$, *then a (finite or infinite) prefix* $u$ *of* $w$, *of length* $|u| > i$, *satisfies* $\Psi$, *and vice-versa (i.e., if a prefix* $u$ *of* $w$ *satisfies* $\Psi$, *then there is an accepting run on* $w$ *with* $r_i \neq \top$ *for all* $i < |u|$).

We can now finish the informal description of the construction. In essence, $\mathsf{A}_\varphi$ guesses the $g$ descendants $y_1, \ldots, y_g$ of the root of the input tree as given in Proposition 8. For every $1 \leq i \leq g$, the automaton uses one copy of $\mathbb{A}^{\neg\Psi}$ to verify that the path $\pi$, from the root to the father of $y_i$, is not $\psi$-conservative (by guessing some finite or infinite extension $\pi \preccurlyeq \pi'$ of it such that $\pi' \models \neg\Psi$), and one copy of $\mathbb{A}_\Psi$ to guess an infinite path $\pi''$ from the root through $y_i$ such that $\pi'' \models \Psi$ (and is thus $\psi$-conservative).

### The construction of GHTA $\mathsf{A}_\vartheta$ for a GCTL* formula $\vartheta$.

We proceed by induction on the structure of $\vartheta$. Given a state sub-formula $\phi$ of $\vartheta$ (possibly including $\vartheta$), for every formula $\theta \in \overline{max(\phi)}$, let $\mathsf{A}_\theta = \langle \Sigma, Q^\theta, q_0^\theta, \delta^\theta, \langle G^\theta, B^\theta \rangle, \langle \mathsf{part}^\theta, type^\theta, \preccurlyeq^\theta \rangle \rangle$ be a GHTA accepting the finitely branching tree models of $\theta$. We build the GHTA $\mathsf{A}_\phi$ accepting all finitely branching tree models of $\phi$ by suitably composing the automata of its maximal sub-formulas and their negations. Note that when composing these automata, we assume w.l.o.g. that the states of any occurrence of a constituent automaton of a sub-formula are disjoint from the states of any other occurrence of a constituent automaton (of the same or of a different sub-formula), as well as from any newly introduced states.[3] Formally, $\mathsf{A}_\phi$ is constructed as follows:

1. If $\phi = p \in AP$, then $\mathsf{A}_\phi = \langle \Sigma, \{q\}, q, \delta, \langle \emptyset, \emptyset \rangle, \langle \mathsf{part}, \mathsf{type}, \preccurlyeq \rangle \rangle$ where $\delta(q, \sigma) = \mathbf{true}$ if $p \in \sigma$ and $\mathbf{false}$ otherwise, $\mathsf{part} = \{\{q\}\}$, $\mathsf{type}(\{q\}) = trans$, and $\preccurlyeq$ is the empty relation.
2. If $\phi = \varphi_0 \vee \varphi_1$ then $\mathsf{A}_\phi$ is obtained by nondeterministically launching either $\mathsf{A}_{\varphi_0}$ or $\mathsf{A}_{\varphi_1}$.
3. If $\phi = \neg\varphi$, then $\mathsf{A}_\phi$ is obtained by dualizing the automaton $\mathsf{A}_\varphi$. Formally, the *dual of a GHTA* $\mathsf{A}$ is the GHTA obtained by dualizing the transition function of $\mathsf{A}$ (i.e., switch $\vee$ and $\wedge$, switch $\top$ and $\bot$, and switch $\square$ and $\lozenge$), replacing the acceptance condition $\langle G, B \rangle$ with $\langle B, G \rangle$, and dualizing the types (i.e., every existential set becomes a universal set, and vice versa).

---

[3] For example, when building an automaton for $\phi = \varphi_0 \vee \varphi_1$ (item 2), in the degenerate case that $\varphi_0 = \varphi_1$ then $\mathsf{A}_{\varphi_1}$ is taken to be a copy of $\mathsf{A}_{\varphi_0}$ with its states renamed to be disjoint from those of $\mathsf{A}_{\varphi_0}$. Also, the new state $q_0$ may be renamed to avoid a collision with any of the other states.

4. If $\phi = \mathsf{E}^{\geq g}\psi$, then $\mathsf{A}_\phi = \langle \Sigma, Q, q_0, \delta, \langle G, B \rangle, \langle \mathsf{part}, \mathsf{type}, \preceq \rangle \rangle$ and its structure is detailed below. Observe that $\psi$ is a path formula and, by Lemma 3, reasoning about $\psi$ can be reduced to reasoning about the LTL formula $\Psi$ whose atoms are elements of $max(\psi)$. Let $\Sigma' = 2^{max(\psi)}$. By Theorem 9, there is an NBW $\mathbb{A}_\Psi = \langle \Sigma', Q^+, q_0^+, \delta^+, G^+ \rangle$ accepting all infinite words in $\Sigma'^\omega$ satisfying $\Psi$. By Lemma 10, there is an NBW $\mathbb{A}^{\neg\Psi} = \langle \Sigma', Q^\neg, q_0^\neg, \delta^\neg, G^\neg \rangle$ accepting all infinite words in $\Sigma'^\omega$ that either satisfy $\neg\Psi$ or have some prefix that satisfies $\neg\Psi$.

   **The set of states.** $Q = Q_1 \cup Q_2$, where $Q_1 = (Q^+ \cup \{\bot\})^g \times (Q^\neg \cup \{\bot\})^g \setminus \{\bot\}^{2g}$, and $Q_2 = \bigcup_{\theta \in \overline{max(\psi)}} Q^\theta$. The $Q_1$ states are used to run $g$ copies of $\mathbb{A}^{\neg\Psi}$ and $g$ copies of $\mathbb{A}_\Psi$ in parallel. Every state in $Q_1$ is a vector of $2g$ coordinates where coordinates $1,...,g$ contain states of $\mathbb{A}_\Psi$, and coordinates $g+1,...,2g$ contain states of $\mathbb{A}^{\neg\Psi}$. In addition, each coordinate may contain the special symbol $\bot$ indicating that it is *disabled*, as opposed to *active*. Note that we disallow the vector $\{\bot\}^{2g}$ that has all coordinates disabled. States in $Q_2$ are all those from the automata $\mathsf{A}_\theta$ for every maximal state subformula of $\psi$, or its negation, and used to run copies of these automata whenever $\mathsf{A}_\phi$ guesses that $\theta$ holds at a given node. We call the coordinates $\{1\ldots g\}$ the $\Psi$ *coordinates*, and the ones $\{g+1\ldots 2g\}$ the $\neg\Psi$ *coordinates*. Also, for every $1 \leq i \leq g$, we denote by $Q_{single}^i = \{(q_1,...,q_{2g}) \in Q_1 \mid q_i \neq \bot$, and for all $j \leq g$, if $j \neq i$ then $q_j = \bot\}$ the set of all states in $Q_1$ in which the only active $\Psi$ coordinate is $i$; and by $Q_{single}^{\leq g} = \cup_{1 \leq i \leq 2g} Q_{single}^i$ the set of all states in $Q_1$ in which exactly one $\Psi$ coordinate is active.

   **The initial state.** $q_0 = (q_1,...,q_{2g})$ where for every $1 \leq i \leq g$ we have that $q_i = q_0^+$ and for every $g+1 \leq i \leq 2g$ we have that $q_i = q_0^\neg$.

   **The acceptance condition.** $B = \cup_{\theta \in \overline{max(\psi)}} B^\theta$ and $G = G' \cup (\cup_{\theta \in \overline{max(\psi)}} G^\theta)$, where $G' = \{(q_1,...,q_{2g}) \in Q_{single}^i \mid q_i \in G^+\}$ is the set of all states in $Q_1$ in which the only active $\Psi$ coordinate contains a good state.

   **The transition function** $\delta$ is defined, for every $\sigma \in \Sigma$, as follows:
   - For every $q \in Q_2$, let $\theta \in \overline{max(\psi)}$ be such that $q \in Q^\theta$, and define $\delta(q, \sigma) = \delta^\theta(q, \sigma)$. Thus, for states in $Q_2$, we simply follow the rules of their respective automata.
   - For every $q \in Q_1$, we define

   $$\delta(q, \sigma) = \bigvee_{\sigma' \in \Sigma'} \left[ (\bigvee_{X \in Legal(q, \sigma')} \Diamond(X)) \bigwedge (\wedge_{\theta \in \sigma'} \delta^\theta(q_0^\theta, \sigma)) \bigwedge (\wedge_{\theta \notin \sigma'} \delta^{\neg\theta}(q_0^{\neg\theta}, \sigma)) \right]$$

   where $Legal(q, \sigma')$ is the set of all *legal distributions* of $(q, \sigma')$, and is defined later. The meaning of this transition is as follows. The disjunction $\bigvee_{\sigma' \in \Sigma'}$ corresponds to all possible guesses of the set of maximal subformulas of $\psi$ that currently hold. Once a guess $\sigma'$ is made, the copies of $\mathbb{A}^{\neg\Psi}$ and $\mathbb{A}_\Psi$ simulated by the states appearing in $Legal(q, \sigma')$ proceed as if the input node was labeled by the letter $\sigma'$. The conjunction $(\wedge_{\theta \in \sigma'} \delta^\theta(q_0^\theta, \sigma)) \wedge (\wedge_{\theta \notin \sigma'} \delta^{\neg\theta}(q_0^{\neg\theta}, \sigma))$ ensures that a guess is correct by launching a copy of $\mathsf{A}_\theta$ for every subformula $\theta \in \sigma'$ that was guessed to hold, and a copy of $\mathsf{A}_{\neg\theta}$ for every subformula $\theta$ guessed not to hold.

   We now define what a *legal distribution* is. Intuitively, a legal distribution of $(q, \sigma')$ is a sequence $q^1,...,q^m$ of different states from $Q_1$ that "distribute" among them, without duplication, the coordinates active in $q$, while making sure that for every $1 \leq i \leq g$ coordinate $i$ (which simulates a copy of $\mathbb{A}_\Psi$) does not get separated from the coordinate $i + g$ (which simulates its partner copy of $\mathbb{A}^{\neg\Psi}$) for as long as $i$ is not the only active $\Psi$ coordinate in the state. As expected, every active coordinate $j$, in any of the states $q^1,...,q^m$, follows from $q_j$ by using the transitions available in the automaton it simulates: $\mathbb{A}_\Psi$ if $j \leq g$, or $\mathbb{A}^{\neg\Psi}$ if $j > g$.

More formally, given a letter $\sigma' \in \Sigma'$, and a state $q = (q_1,...q_{2g}) \in Q_1$ in which the active coordinates are $\{i_1,...,i_k\}$, we say that a sequence $X = q^1,...,q^m$ (for some $m \geq 1$) of distinct states in $Q_1$ is a *legal distribution* of $(q, \sigma')$ if the following conditions hold: *(i)* the coordinates active in the states $q^1,...,q^m$ are exactly $i_1,...,i_k$, i.e., $\{i_1,...,i_k\} = \cup\{i \in \{1,...,2g\} \mid \exists 1 \leq l \leq m \text{ s.t. } q_i^l \neq \bot\}$. *(ii)* if a coordinate $i_j$ is active in some $q' \in X$ then it is not active in any other $q'' \in X$; *(iii)* if $1 \leq i_j < i_l \leq g$ are two active $\Psi$ coordinates in some $q' \in X$, then $q'_{i_j+g}, q'_{i_l+g} \in Q^\neg \setminus \{\top\}$, i.e., the coordinates $i_j + g, i_l + g$ are also active in $q'$ and do not contain the accepting sink of $\mathbb{A}^{\neg\Psi}$; *(iv)* if $i_j$ is active in some $q' \in X$ then $(q_{i_j}, \sigma', q'_{i_j}) \in \delta^+$ if $i_j \leq g$, and $(q_{i_j}, \sigma', q'_{i_j}) \in \delta^\neg$ if $i_j > g$. I.e., active $\Psi$ coordinates evolve according to the transitions of $\mathbb{A}_\Psi$, and active $\neg\Psi$ coordinates according to the transitions of $\mathbb{A}^{\neg\Psi}$.

**The hesitant structure** Given a set of coordinates $I \subseteq \{1,...,2g\}$, let $\mathbb{Q}_I \subseteq Q_1$ be the set of all vectors whose active coordinates are exactly $I$. We set $\mathsf{type}(\mathbb{Q}_I) = exist$, and set the partitioning $\mathsf{part}$ of $Q$ be the union of $\cup_{I \subseteq \{1,...,2g\}}\{\mathbb{Q}_I\}$ and $\cup_{\theta \in \overline{max(\psi)}}(\mathsf{part}^\theta)$. For every $I \subseteq \{1,...,2g\}$, we have $\mathbb{Q}_J \prec \mathbb{Q}_I$ for every $J \subset I$, and $\mathbb{Q} \preceq \mathbb{Q}_I$ for every $\mathbb{Q} \in \cup_{\theta \in \overline{max(\psi)}}(\mathsf{part}^\theta)$. Observe that if a transition $\delta(q, \sigma)$ from some state $q \in \mathbb{Q}_I$, on some letter $\sigma$, refers to another state $q' \in \mathbb{Q}_I$ then $q$ was not split (since $q'$ has the same active coordinates as $q$), i.e., the $\diamond$ in which $q'$ occurs is of the form $\diamond(q')$. Hence, by the definition of $\delta(q, \sigma)$, there is no other $q'' \in \mathbb{Q}_I$ that is conjuncted with $q'$ in $expand_d(\delta(q, \sigma))$ for any $d$, and thus $\delta(q, \sigma)$ respects the hesitancy constraint.

This completes the formal definition of the construction of the automaton $\mathsf{A}_\vartheta$. The interested reader can find a formal proof of correctness (i.e., the GHTA $\mathsf{A}_\vartheta$ accepts all the finitely-branching tree models of $\vartheta$) in Appendix C. Together with the following straightforward analysis, Theorem 7 is proved:

▶ **Proposition 11.** *The automaton $\mathsf{A}_\vartheta$ is a GHTA, its depth is $O(|\vartheta|)$, it has $2^{O(|\vartheta| \times \mathsf{deg}(\vartheta))}$ many states, and the size of its transition function is $2^{O(|\vartheta| \times \mathsf{deg}(\vartheta))}$.*

▶ **Remark.** We end the section with two observations. First, the $2g$ copies of $\mathbb{A}^{\neg\Psi}$ and $\mathbb{A}_\Psi$ can not simply be launched from the root of the tree using a conjunction in the transition relation. The reason is that if this is done then there is no way to enforce property (i) of Proposition 8. Second, a cursory look may suggest that different copies of $\mathbb{A}^{\neg\Psi}$ and $\mathbb{A}_\Psi$ that are active in the current vector may be merged. Unfortunately, this cannot be done since $\mathbb{A}^{\neg\Psi}$ and $\mathbb{A}_\Psi$ are nondeterministic, and thus, different copies of these automata must be able to make independent guesses in the present in order to accept different paths in the future.

## 5 Complexity of Satisfiability and Model-checking of GCTL*

A first consequence of the previous section is a boundedly-branching tree model property:

▶ **Theorem 12.** *If a GCTL* formula $\vartheta$ is satisfiable then it has a tree model of branching degree at most $2^{O(|\vartheta| \times \mathsf{deg}(\vartheta))}$.*

**Sketch.** Suppose $\vartheta$ is satisfiable. By Theorem 4, $\vartheta$ has a finitely-branching tree model. We prove that every tree model has a subtree of branching degree $|Q|^2$ that is also a model of $\vartheta$, where $Q$ is the state set of the automaton $\mathsf{A}_\vartheta$. Since by Theorem 7 we have that $|Q| = 2^{O(|\vartheta| \times \mathsf{deg}(\vartheta))}$, the theorem follows. The main idea is to build the membership game $G_{\mathsf{T},\mathsf{A}_\vartheta}$ of tree $\mathsf{T}$ and the automaton $\mathsf{A}_\vartheta$. The game is played by two players, *automaton* and *pathfinder*. Player automaton moves by resolving disjunctions, and is trying to show that $\mathsf{T}$ is accepted by $\mathsf{A}_\vartheta$. Player pathfinder moves by resolving conjunctions and is trying to show that $\mathsf{T}$ is not accepted by $\mathsf{A}_\vartheta$. The game uses auxilliary tree arenas to resolve the transition function, a simple case of what [1] calls *hierarchical parity games*. As usual, player automaton has a winning strategy if and only if $\mathsf{T} \models \mathsf{A}_\vartheta$. By memoryless determinacy of parity games on infinite arenas, player automaton has a winning strategy if and only if he has a memoryless winning strategy. For a fixed memoryless strategy $str$, one can prove, by looking at the transition function of $\mathsf{A}_\vartheta$, that every play consistent with $str$, and every node $t$ of the input tree $\mathsf{T}$, only visits at most $|Q|^2$ sons of $t$, thus inducing the required subtree. ◀

▶ **Theorem 13.** *The satisfiability problem for* GCTL$^*$ *over LTS is 2*ExpTime-Complete.

**Proof.** Theorems 7, 12 and Lemma 6, yield that satisfiability of GCTL$^*$ is solvable in time double exponential in $O(|\vartheta| \times \deg(\vartheta))$. The lower bound already holds for CTL$^*$ [23]. ◀

▶ **Theorem 14.** *Model checking* GCTL$^*$ *for finite LTSs is* PSpace-Complete.

## 6 Discussion

This work shows that adding graded path modalities to CTL$^*$ (i.e., GCTL$^*$) results in an expressive logic (it is equivalent over trees to Monadic Path Logic), whose satisfiability and model-checking problems have the same complexity as that of CTL$^*$. Thus our work follows the important tradition of extending logics with graded modalities at no extra cost in complexity of fundamental algorithmic problems. Moreover, GCTL$^*$ inherits all the advantages of CTL$^*$ over other logics such as CTL and $\mu$-calculus (e.g., it can express fairness and is user friendly).

The exact complexity of GCTL$^*$ was conjectured in [4] to be 2ExpTime-Complete. However, as the authors note in the conference version of that paper, their techniques, that worked for GCTL, do not work for GCTL$^*$. Moreover, they also suggested a line of attack that does not seem to work; indeed, it was left out of the journal version of their paper [5]. Instead, our method was a careful combination of the automata-theoretic approach to branching-time logics [17], a novel characterization of the graded path modality (Proposition 8), and a boundedly-branching tree model property whose proof uses game-theoretic arguments (Theorem 12). Moreover, our technique immediately specializes to recover the main results about GCTL from [4], i.e., satisfiability for GCTL is ExpTime-Complete and the model checking problem for GCTL and finite LTS is in PTime.

When investigating the complexity of a logic with a form of counting quantifiers, one must decide how the numbers in these quantifiers contribute to the length of a formula, i.e., to the input of a decision procedure. In this paper we assume that these numbers are coded in unary, rather than binary. There are a few reasons for this. First, the unary coding naturally appears in several practical applications of graded modalities and in particular in related works about description logics [7]. Second, although the complexity of the binary case is often the same as that of the unary case [6, 5], the constructions are significantly more complicated and so less easy to implement. At any rate, as the binary case is useful in some circumstances, we also plan to investigate this.

Now that the exact complexity of GCTL$^*$ has been determined, we suggest several directions for future work. First, recall that logics extended with graded world modalities have been further enriched with backward-modalities and with nominals [6]. We suggest that a similar direction should be taken for graded path modalities, and GCTL* in particular. Second, recall that the graded $\mu$-calculus was used to solve questions (such as satisfiability) for the description logic $\mu\mathcal{ALCQ}$ [6]. Similarly, our techniques for GCTL* might be useful for solving questions in $\mathcal{ALCQ}$ combined with temporal logic, such as for the graded extension of CTL*$_{ALC}$ [13]. Finally, the GCTL model checking algorithm from [5] has been implemented in the NuSMV tool providing the extra ability of finding more counter-examples (if there are any) in one shot while checking a system against a CTL formula. We suggest that the same should be done for GCTL*.

## A Appendix to Section 2: The GCTL* temporal logic

### A.1 Notation of standard objects

A *Labelled Transition System (LTS)* is written $\mathsf{S} = \langle \Sigma, S, E, \lambda \rangle$ where $\Sigma$ is a set of *labels*, $S$ is a finite or countably infinite set of *states*, $E \subseteq S \times S$ is the *transition relation*, and $\lambda : S \mapsto \Sigma$ is the *labeling function*. Typically $\Sigma = 2^{\mathrm{AP}}$ where AP is a set of *atoms* and $\lambda(q)$ is a finite subset of AP. The *degree* of a state $s$ is the cardinality of the set $\{t \in S : (s,t) \in E\}$. For simplicity of presentation we assume that $E$ is total, i.e., for every $q \in S$ there exists $q' \in S$ such that $(q, q') \in E$.

A path in $\mathsf{S}$ is a **finite or infinite** sequence $\pi_0 \pi_1 \cdots \in (S^*) \cup (S^\omega)$ such that $(\pi_i, \pi_{i+1}) \in E$ for all $i < |\pi|$. Here $|\pi|$, the *length* of the path $\pi$, is defined to be the cardinality of the sequence $\pi$, and in particular it is equal to $\omega$ if the path is infinite. Note that we count positions in the sequence starting with 0. If $0 \leq i < |\pi|$ then write $\pi_{\geq i}$ for the suffix of $\pi$ starting at position $i$, namely the path $\pi_i \pi_{i+1} \cdots \in (S^*) \cup (S^\omega)$. The set of paths in $\mathsf{S}$ is written $\mathsf{pth}(\mathsf{S})$, and the set of paths in $\mathsf{S}$ that start with $q$ is written $\mathsf{pth}(\mathsf{S}, q)$. Write $\preceq$ for the prefix ordering on paths, and if $\pi \preceq \pi'$ say that $\pi'$ is an *extension* of $\pi$. Note that $\preceq$ is defined on all sets of the form $\mathsf{pth}(\mathsf{S}, q)$. Note that if $\pi$ is infinite then $\pi \preceq \pi'$ implies $\pi = \pi'$. If $X \subseteq \mathsf{pth}(\mathsf{S}, q)$ is a set of paths with the same starting element, define $\min(X)$ as the set of minimal elements in $X$ with respect to the prefix ordering $\preceq$.

A $\Sigma$-*labeled tree* $\mathsf{T}$ is a pair $\langle T, V \rangle$ where $T \subseteq \mathbb{N}^*$ is a $\prec$-downward closed set of strings over $\mathbb{N}$, and $V : T \to \Sigma$ is a labeling, where $\Sigma$ is a set of *labels*. Here $\prec$ is the prefix ordering on $\mathbb{N}^*$. In particular the empty string $\epsilon$ is the root of the tree. For $t \in T$ write $sons(t) \subset T$ for the set $\{s \in T : t \prec s \wedge \neg \exists z . t \prec z \prec s\}$. The *degree of a node* $v$ is the cardinality of $sons(v)$. If every node $v$ of tree $T$ has finite degree, then we say that $T$ is *finitely branching*. If there is a $k \in \mathbb{N}$ such that every node of $v$ of $T$ has degree at most $k$, then say that $T$ is *boundedly branching* or *has branching degree* $k$. We implicitly view a tree $\mathsf{T} = \langle T, V \rangle$ as the LTS $\langle \Sigma, T, E, V \rangle$ where $(t, s) \in E$ iff $s \in sons(t)$. As for LTSs, we assume for simplicity of presentation that trees are total (i.e., $sons(t) \neq \emptyset$ for every $t \in T$).

If $\psi$ is an LTL formula, then we may write $\pi \models \psi$ instead of $(\mathsf{S}, \pi) \models \psi$ (this notation is justifiable since the truth of an LTL formula $\psi$ depends only on the path $\pi$ and not on the rest of the structure $\mathsf{S}$).

Two state formulas $\phi, \phi'$ are *equivalent* if for all $\mathsf{S}$ and $s \in S$, we have $(\mathsf{S}, s) \models \phi$ iff $(\mathsf{S}, s) \models \phi'$. Similarly, two path formulas $\psi, \psi'$ are *equivalent* if for all $\mathsf{S}$ and $\pi \in \mathsf{pth}(\mathsf{S})$, we have that $(\mathsf{S}, \pi) \models \psi$ if and only if $(\mathsf{S}, \pi) \models \psi'$.

### A.2 Syntax and Semantics of GCTL*

▶ **Definition 15** (Syntax of GCTL*)**.** Fix a set of atoms AP. The GCTL* *state ($\varphi$) and path ($\psi$) formulas* are built inductively from AP using the following grammar:

$$\varphi ::= p \mid \neg \varphi \mid \varphi \vee \varphi \mid \mathsf{E}^{\geq g} \psi \qquad \psi ::= \varphi \mid \neg \psi \mid \psi \vee \psi \mid \mathsf{X}\psi \mid \psi \mathsf{U} \psi \mid \psi \mathsf{R} \psi$$

In the first item, $p$ varies over AP and $g$ varies over $\mathbb{N}$ (and thus, technically, there are infinitely many rules in this grammar). As usual, $\mathsf{X}, \mathsf{U}$ and $\mathsf{R}$ are called *temporal operators* and $\mathsf{E}^{\geq g}$ (for $g \in \mathbb{N}$) are called *path quantifiers*. The class of GCTL* *formulas* is the set of state formulas generated by the above grammar. The simpler class of *Graded* CTL *formulas* (GCTL) is obtained by requiring each temporal operator occurring in a formula to be immediately preceeded by a path quantifier, as in the classical definition of CTL. The class of LTL formulas is obtained as the path formulas in which no path quantifier appears.

The *degree* of the quantifier $\mathsf{E}^{\geq g}$ is the number $g$. For a state formula $\varphi$, we define the *degree* $\deg(\varphi)$ of $\varphi$ as the maximum natural number $g$ occurring among the degrees of all its path quantifiers. The *length* of a formula $\varphi$, denoted by $|\varphi|$, is defined inductively on the structure of $\varphi$ as usual, and using $|\mathsf{E}^{\geq g}\psi|$ equal to $g + 1 + |\psi|$. Clearly, $\deg(\varphi) = \mathcal{O}(|\varphi|)$.

▶ **Definition 16** (Semantics of GCTL* and $\psi$-conservative). Fix LTS S. If $\varphi$ is a GCTL* state formula and $s \in S$, then define $(\mathsf{S}, s) \models \varphi$ inductively:

1. $(\mathsf{S}, s) \models p$, for $p \in \mathrm{AP}$, iff $p \in \lambda(s)$.
2. $(\mathsf{S}, s) \models \neg\varphi$ iff $(\mathsf{S}, s) \not\models \varphi$.
3. $(\mathsf{S}, s) \models \mathsf{E}^{\geq g}\psi$, for $\psi$ a GCTL* path formula, iff the cardinality of the set $\min(Con(\mathsf{S}, s, \psi))$ is at least $g$, where $Con(\mathsf{S}, s, \psi) := \{\pi \in \mathsf{pth}(\mathsf{S}, s) \mid \forall \pi' \in \mathsf{pth}(\mathsf{S}, s) : \pi \preceq \pi'$ implies $(\mathsf{S}, \pi') \models \psi\}$. The paths in $Con(\mathsf{S}, s, \psi)$ are called $\psi$-*conservative (in* S *starting at* $s$*)*, and paths in $\min(Con(\mathsf{S}, s, \psi))$ are called *minimal* $\psi$-*conservative*.

If $\psi$ is a GCTL* path formula and $\pi = \pi_0\pi_1\cdots \in \mathsf{pth}(\mathsf{S})$ is a finite or infinite path in S, then define $(\mathsf{S}, \pi) \models \psi$ inductively:

1. $(\mathsf{S}, \pi) \models \varphi$, for $\varphi$ a state formula, iff $(\mathsf{S}, \pi_0) \models \varphi$.
2. $(\mathsf{S}, \pi) \models \neg\psi$ iff $(\mathsf{S}, \pi) \not\models \psi$.
3. $(\mathsf{S}, \pi) \models \psi_1 \vee \psi_2$ iff $(\mathsf{S}, \pi) \models \psi_1$ or $(\mathsf{S}, \pi) \models \psi_2$.
4. $(\mathsf{S}, \pi) \models \mathsf{X}\psi$ iff $|\pi| \geq 2$ and $(\mathsf{S}, \pi_{\geq 1}) \models \psi$.
5. $(\mathsf{S}, \pi) \models \psi_1 \mathsf{U} \psi_2$ iff there exists $i$ with $0 \leq i < |\pi|$ such that $(\mathsf{S}, \pi_{\geq i}) \models \psi_2$, and for all $j$ with $0 \leq j < i$, $(\mathsf{S}, \pi_{\geq j}) \models \psi_1$.
6. $(\mathsf{S}, \pi) \models \psi_1 \mathsf{R} \psi_2$ iff i) for all $i$ with $0 \leq i < |\pi|$, either $(\mathsf{S}, \pi_{\geq i}) \models \psi_2$ or there exists $j$ with $0 \leq j < i$ such that $(\mathsf{S}, \pi_{\geq j}) \models \psi_1$; and morever ii) if $\pi$ is finite then there is some $j < |\pi|$ such that $(\mathsf{S}, \pi_{\geq j}) \models \psi_1$.

An LTS S with a designated state $q \in S$ is a *model* of a GCTL* formula $\varphi$, sometimes denoted $\mathsf{S} \models \varphi$, if $(\mathsf{S}, q) \models \varphi$. For instance, $\mathsf{T} \models \varphi$ means that $(\mathsf{T}, \epsilon) \models \varphi$ where $\epsilon$ is the root of T. A GCTL* formula $\varphi$ is *satisfiable* iff there exists a model of it.

## A.3 Proof of Lemma 3, Page 4

**Proof.** Let $\Psi$ be the LTL formula over atoms $max(\psi)$ corresponding to $\psi$, as defined above. However for the duration of this inductive proof, instead of $\Psi$, write $\widehat{\psi}$. Proceed by induction on $\psi$ (for all $\mathsf{S}, \pi$).

The base case is that $\psi$ is a state formula. Then $\widehat{\psi} \in max(\psi)$. Thus $(\mathsf{S}, \pi) \models \psi$ iff $(\mathsf{S}, \pi_0) \models \psi$ iff $\widehat{\psi}$ (an atom) is in $L(q)$ iff $(\mathsf{S}_\psi, \pi_0) \models \widehat{\psi}$ iff $(\mathsf{S}_\psi, \pi) \models \widehat{\psi}$.

So suppose now that $\psi$ is a path formula that is not a state formula. We have the following cases.

1. Suppose $\psi = \neg\gamma$. Then $max(\psi) = max(\gamma)$, and so $\widehat{\psi} = \neg\widehat{\gamma}$, and so

$$(\mathsf{S}, \pi) \models \psi \text{ iff } (\mathsf{S}, \pi) \not\models \gamma$$
$$\text{iff } (\mathsf{S}_\gamma, \pi) \not\models \widehat{\gamma}$$
$$\text{iff } (\mathsf{S}_{\neg\gamma}, \pi) \not\models \widehat{\gamma}$$
$$\text{iff } (\mathsf{S}_{\neg\gamma}, \pi) \models \neg\widehat{\gamma}$$
$$\text{iff } (\mathsf{S}_{\neg\gamma}, \pi) \models \widehat{\psi}.$$

2. Suppose $\psi = \mathsf{X}\gamma$. Then, $max(\psi) = max(\gamma)$, and so $\widehat{\psi} = \mathsf{X}\widehat{\gamma}$, and so

   $(\mathsf{S}, \pi) \models \psi$ iff $(\mathsf{S}, \pi_{\geq 1}) \models \gamma$ and $|\pi| > 1$
   
   iff $(\mathsf{S}_\gamma, \pi_{\geq 1}) \models \widehat{\gamma}$ and $|\pi| > 1$
   
   iff $(\mathsf{S}_{\mathsf{X}\gamma}, \pi) \models \widehat{\gamma}$ and $|\pi| > 1$
   
   iff $(\mathsf{S}_{\mathsf{X}\gamma}, \pi) \models \mathsf{X}\widehat{\gamma}$
   
   iff $(\mathsf{S}_{\mathsf{X}\gamma}, \pi) \models \widehat{\psi}$.

3. Suppose $\psi = \gamma \mathsf{U} \zeta$. Then, $max(\psi) = max(\gamma) \cup max(\zeta)$, and so $\widehat{\psi} = \widehat{\gamma} \mathsf{U} \widehat{\zeta}$. Then $(\mathsf{S}, \pi) \models \psi$ iff

   $\exists i. 0 \leq i < |\pi|.(\mathsf{S}, \pi_{\geq i}) \models \zeta$, and $\forall j. 0 \leq j < i, (\mathsf{S}, \pi_{\geq j}) \models \gamma$ iff
   
   $\exists i. 0 \leq i < |\pi|.(\mathsf{S}_\zeta, \pi_{\geq i}) \models \widehat{\zeta}$, and $\forall j. 0 \leq j < i, (\mathsf{S}_\gamma, \pi_{\geq j}) \models \widehat{\gamma}$ iff
   
   $\exists i. 0 \leq i < |\pi|.(\mathsf{S}_\psi, \pi_{\geq i}) \models \widehat{\zeta}$, and $\forall j. 0 \leq j < i, (\mathsf{S}_\psi, \pi_{\geq j}) \models \widehat{\gamma}$,

   which is equivalent to $(\mathsf{S}_\psi, \pi) \models \widehat{\gamma} \mathsf{U} \widehat{\zeta}$, which is equivalent to $(\mathsf{S}_\psi, \pi) \models \widehat{\psi}$.

   The other cases, $\vee$ and $\mathsf{R}$, are similar to the $\mathsf{U}$ case. ◂

## A.4 Proof of Theorem 4, Page 4

**Proof of Theorem 4 Part 1.** The *unwinding* of LTS $\mathsf{S} = \langle 2^{\mathrm{AP}}, S, E, \lambda \rangle$ and state $t \in S$ is the LTS $\mathsf{S}^t = \langle 2^{\mathrm{AP}}, S', E', \lambda' \rangle$ where $S'$ is the set of finite non-empty paths of $\mathsf{S}$ starting in $t$, $(\pi, \rho) \in E'$ iff $\rho = \pi r$ for some $r \in S$, and $\lambda'(\pi r) := \lambda(r)$, for $\pi \in S' \cup \{\epsilon\}$ (here $\epsilon$ is the empty string, and $\epsilon r = r$). Note that the unwinding of an LTS is a $2^{\mathrm{AP}}$-labelled tree.

Notation. We use '·' for concatenation of paths in $\mathsf{S}^t$, and we use adjacency or '.' for concatenation of paths in $\mathsf{S}$.

The statement we want to prove is this: for all GCTL* formulas $\phi$ (these are state formulas), and all LTSs $\mathsf{S}$, and all $t \in S$, $(\mathsf{S}, t) \models \phi$ if and only if $(\mathsf{S}^t, t) \models \phi$.

We first define a natural bijection $f$ between $\mathsf{pth}(\mathsf{S}, t)$ and $\mathsf{pth}(\mathsf{S}^t, t)$.

Define $f(t) := t$, and $f(\pi_0 \ldots \pi_n r) := f(\pi_0 \ldots \pi_n) \cdot \pi_0 \ldots \pi_n r$, and $f(\pi_0 \pi_1 \ldots) := \pi_0 \cdot \pi_0 \pi_1 \cdot \pi_0 \pi_1 \pi_2 \cdot \ldots$.

Note that $f$ is $\preceq$-preserving, i.e., $\pi \preceq \pi'$ iff $f(\pi) \preceq f(\pi')$.

The inductive hypothesis for a GCTL* formula $\alpha$ says that for all LTSs $\mathsf{S}$, $t \in S$, and $\pi \in \mathsf{pth}(\mathsf{S}, t)$:

- if $\alpha$ is a state formula then $(\mathsf{S}, t) \models \alpha$ iff $(\mathsf{S}^t, t) \models \alpha$,
- if $\alpha$ is a path formula then $(\mathsf{S}, \pi) \models \alpha$ iff $(\mathsf{S}^t, f(\pi)) \models \alpha$.

Suppose the inductive hypothesis holds for all proper subformulas of $\alpha$. Fix $\mathsf{S}, t, \pi$. There are two main cases.

**Suppose $\alpha$ is a state formula.** There are three cases.

1. Suppose $\alpha$ is of the form $p$ for $p \in \mathrm{AP}$. In this case we must prove that $(\mathsf{S}, t) \models p$ iff $(\mathsf{S}^t, t) \models p$, which is immediate from the definition of $\lambda'$.
2. The case that $\alpha = \neg \varphi_1$ or $\alpha = \varphi_1 \vee \varphi_2$ for state formulas $\varphi_i$ is immediate from the semantics of these Boolean operations and the inductive hypothesis.
3. Suppose $\alpha$ is of the form $\mathsf{E}^{\geq g} \psi$ for path formula $\psi$.
   For the first direction, suppose $(\mathsf{S}, t) \models \alpha$, i.e., there are at least $g$ many minimal $\psi$-conservative paths in $\mathsf{pth}(\mathsf{S}, t)$. In other words, there exists distinct $\pi_1, \cdots, \pi_g \in \mathsf{pth}(\mathsf{S}, t)$ such that for every $i$, we have

  **a.** Every extension $\pi' \in \mathsf{pth}(\mathsf{S},t)$ of $\pi_i$ satisfies $(\mathsf{S},\pi') \models \psi$.
  **b.** Every prefix $\tau$ of $\pi_i$ has an extension $\rho \in \mathsf{pth}(\mathsf{S},t)$ that satisfies $(\mathsf{S},\rho) \not\models \psi$.

Thus: $f(\pi_1), \cdots, f(\pi_g) \in \mathsf{pth}(\mathsf{S}^t, t)$ are distinct, and for every $i$ we have:

  **a.** Every extension $\pi' \in \mathsf{pth}(\mathsf{S}^t, t)$ of $f(\pi_i)$ satisfies $(\mathsf{S}^t, \pi') \models \psi$. To see this note that $\pi' = f(\pi)$ for some $\pi \in \mathsf{pth}(\mathsf{S}, t)$, and so $f(\pi_i) \preceq f(\pi)$, and so $\pi_i \preceq \pi$, and so by 3a. $(\mathsf{S}, \pi) \models \psi$, and so by induction $(\mathsf{S}^t, f(\pi)) \models \psi$.
  **b.** Every prefix $\tau'$ of $f(\pi_i)$ has an extension $\rho' \in \mathsf{pth}(\mathsf{S}^t, t)$ that satisfies $(\mathsf{S}^t, \rho') \not\models \psi$ (use similar reasoning to the previous case).

Thus $(\mathsf{S}^t, t) \models \mathsf{E}^{\geq g}\psi$, and this completes the first direction. The other direction, i.e., that $(\mathsf{S}^t, t) \models \mathsf{E}^{\geq g}\psi$ implies $(\mathsf{S}, t) \models \mathsf{E}^{\geq g}\psi$ is done by simply reversing the argument for the first direction.

**Suppose $\alpha$ is a path formula, say $\psi$.** Let $\Psi$ be the LTL formula from Lemma 3. Then

$$(S, \pi) \models \psi \text{ iff } (S_\psi, \pi) \models \Psi$$
$$\text{iff } ((S_\psi)^t, f(\pi)) \models \Psi$$
$$\text{iff } ((S^t)_\psi, f(\pi)) \models \Psi$$
$$\text{iff } (S^t, f(\pi)) \models \psi.$$

The first and fourth equivalences follow from Lemma 3, the second and third equivalences follows from inductive hypothesis applied to the maximal state sub-formulas of $\psi$ and the fact that $\pi \in S_\psi$ and $f(\pi) \in (S_\psi)^t$ and $f(\pi) \in (S^t)_\psi$ induce the same infinite sequence of labels (and thus the paths agree on the LTL formula $\Psi$). ◂

**Proof of Theorem 4 Part 2.** We briefly summarise the syntax and semantics of MPL (or see [19, Section 1.2]). For a tree $\mathsf{T}$ write $\mathsf{branches}(\mathsf{T})$, the *branches of* $\mathsf{T}$, for those finite or infinite paths of $\mathsf{T}$, starting from the root, that are maximal (i.e., have no proper extensions in $\mathsf{T}$). The syntax of MPL has logical symbols for the Boolean operations, first-order variables $x, y, \cdots$, path variables $X, Y, \cdots$, quantification over these variables, and non-logical symbols $=, \prec, \epsilon$ and $L_p$ for atoms $p \in \mathsf{AP}$. The semantics are defined for labeled trees $\mathsf{T} = \langle T, V \rangle$ where $V : T \to 2^{\mathsf{AP}}$. The interpretation of variables $x$ are over elements of $T$, the interpretation of variables $X$ are over branches of $T$, the interpretation of $=$ is the usual equality of variables, the interpretation of $\prec$ is as the ancestor relation of $T$, the interpretation of $x \in X$ is that node $x$ is on the branch $X$, and the interpretation of $L_p$ is as the set of nodes $t$ of $T$ such that $p \in V(t)$.

In this proof we freely switch between viewing $\mathsf{T}$ as a tree and as a LTS.

To show that MPL is at least as expressive as $\mathsf{GCTL}^*$ over trees, we show: (†) for every $\mathsf{GCTL}^*$ state formula $\phi$ there exists an MPL formula $\widehat{\phi}(x)$ such that for all trees $\mathsf{T}$, and all $t \in \mathsf{T}$: $(\mathsf{T}, t) \models \phi$ if and only if $\mathsf{T} \models \widehat{\phi}(t)$.

We start with some notation and two lemmas.

Notation. For $\pi \in \mathsf{pth}(\mathsf{T})$ and $a \in \pi$ write $\pi_{[a,\infty)} \in \mathsf{pth}(\mathsf{T}, a)$ for the tail of $\pi$ starting at $a$. Also, for $a, b \in T$ with $a \preceq b$, write $\pi_{[a,b]} \in \mathsf{pth}(\mathsf{T}, a)$ for the subpath of $\pi$ starting at $a$ and ending at $b$.

Fact 1. For every LTL formula $\Psi$ over atoms AP there is an MPL formula $\Psi'(x, X)$ (whose atomic relations are of the form $L_p$ for $p \in \mathsf{AP}$) such that for all trees $\mathsf{T}$, and all $a \in T$ and all paths $\pi \in \mathsf{branches}(\mathsf{T})$ with $a \in \pi$: $T \models \Psi'(a, \pi)$ if and only if $(T, \pi_{[a,\infty)}) \models \Psi$.

Fact 2. For every LTL formula $\Psi$ over atoms AP there is an MPL formula $\Psi''(x, y)$ (whose atomic relations are of the form $L_p$ for $p \in$ AP) such that for all trees T, and all $a, b \in T$ with $a \preceq b$: $T \models \Psi''(a, b)$ if and only if $(T, \pi_{[a,b]}) \models \Psi$.

We prove Fact 2 (the proof of Fact 1 is similar). Construct $\Psi''(x, y)$ by induction on the formula $\Psi$. If $\Psi$ is an atom, say $p \in$ AP, then $\Psi''(x, y)$ is defined as $L_p(x)$. If $\Psi = \neg \Psi_1$ then $\Psi''(x, y)$ is defined as $\neg \Psi_1''(x, y)$. Similarly for the case that $\Psi$ is a disjunction. If $\Psi = \Psi_1 \mathsf{U} \Psi_2$ then $\Psi''(x, y)$ is defined as $\exists z. x \preceq z \preceq y \wedge [\Psi_2''(x, y) \wedge \forall v. x \preceq v \prec z \rightarrow \Psi_1'(x, v)]$. The cases X and R are similar to U. This completes the proof of Fact 2.

Fact 3. For every LTL formula $\Psi$ over atoms AP there is an MPL formula $mincon_\Psi(x, X)$ (whose atomic relations are of the form $L_p$ for $p \in$ AP) such that for all trees T, and all $a \in T$ and all branches $\pi \in T$ with $a \in \pi$: $T \models mincon_\Psi(a, \pi)$ if and only if the tail of $\pi$ starting at $a$ is minimal $\Psi$-conservative in $(T, a)$. Similarly, there is a formula $mincon_\Psi(x, y)$ such that for all trees T and all $a, b \in T$ with $a \preceq b$: $T \models mincon_\Psi(a, b)$ if and only if the path between $a$ and $b$ is minimal $\Psi$-conservative in $(T, a)$.

Proof of Fact 3.
- Let $end(X, z)$ denote the formula $z \in X \wedge \forall y \in X. y \preceq z$ stating that $z$ is the last node of the branch $X$.
- Let $finite(X)$ denote the formula $\exists z. end(X, z)$ stating that branch $X$ is finite.
- We use the shorthand $end(X)$ for the unique value $end(X, z)$ if it exists.
- Let $finmincon_\psi(X, x)$ denote the formula

$$finite(X) \wedge (\forall y. end(X) \preceq y \rightarrow \Psi''(x, y))$$
$$\wedge (\forall Y. end(X) \in Y \rightarrow \Psi'(x, Y))$$
$$\wedge (\forall z. x \preceq z \prec end(X) \rightarrow \neg \Psi''(x, z))$$

   stating that the path $X$ starting at $x$ is finite and minimal $\Psi$-conservative.
- Let $infmincon_\psi(X, x)$ denote the formula

$$\neg finite(X) \wedge \Psi'(x, X) \wedge (\forall z. x \preceq z \in X \rightarrow \neg \Psi''(x, z))$$

   stating that the path $X$ starting at $x$ is infinite and minimal $\Psi$-conservative.
- Finally, let $mincon_\Psi(x, X)$ denote the formula $finmincon_\psi(X, x) \vee infmincon_\psi(X, x)$ stating that the path $X$ starting at $x$ is minimal $\Psi$-conservative (irrespective of $X$ being finite or infinite).

Similarly, the formula $mincon_\Psi(x, w)$ is defined as in $finmincon_\psi(X, x)$ but replacing $end(X)$ by $w$.

This completes the proof of Fact 3.

We now show how to inductively define the formula $\widehat{\phi}(x)$ in (†):
- If $\phi$ is an atom, say $p$, then $\widehat{\phi}(x)$ is defined as the unary predicate $L_p(x)$.
- If $\phi$ is of the form $\neg \phi'$, then $\widehat{\phi}$ is defined as $\neg \widehat{\phi}$; and similarly for $\vee$ and $\wedge$.
- If $\phi$ is of the form $\mathsf{E}^{\geq g} \psi$, then let $\Psi$ be the LTL formula corresponding to $\psi$ from Lemma 3 over atoms $max(\psi)$. For each atom $\theta \in max(\psi)$, let $\hat{\theta}(x)$ be the corresponding MPL formula (which exists by induction) whose atoms are of the form $L_\theta$ for $\theta \in max(\psi)$. The formula $\widehat{\phi}(x)$ is defined as:

$$\bigvee_{h \in [0,g]} \exists X_1, \cdots, X_h. \exists x_1, \cdots, x_{g-h}. [\wedge_{i \in [1,h]} mincon_\Psi(x, X_i)[L_\theta(z)/\hat{\theta}(z)] \bigwedge$$

$$[\wedge_{i \in [1,g-h]} mincon_\Psi(x, x_i)[L_\theta(z)/\hat{\theta}(z)]],$$

where $mincon_\Psi(x, X_i)[L_\theta(z)/\hat{\theta}(z)]$ is the MPL formula $mincon_\Psi(x, X_i)$ in which every occurence of a subformula of the form $L_\theta(z)$ (for $\theta \in max(\psi)$ and $z$ a variable) is replaced by the formula $\hat{\theta}(z)$.

This completes the proof that MPL is at least as expressive as GCTL$^*$. For the other direction we use an intermediate logic, counting-CTL$^*$. Following [19, Section 1.3], we define counting-CTL$^*$ by adding to CTL$^*$ the state formulas $\mathsf{D}^n\phi$, for state formula $\phi$ and $n \in \mathbb{N}$, which are interpreted as saying that at least $n$ children of the current node satisfy $\phi$. By Remark 3, GCTL$^*$ is at least as expressive as counting-CTL$^*$. The main result of [19] is that counting-CTL$^*$ is expressively equivalent to MPL over trees. Thus GCTL$^*$ and MPL are expressively equivalent over trees. ◀

**Proof of Theorem 4 Part 3.** There are a number of ways to prove this result, i.e., that GCTL$^*$ has the finitely-branching tree model property. Although this result follows from the more general statement for full MSO (in which one can embed GCTL$^*$), a result that seems to be folklore and follows from [25, Lemma 43], we also give a proof that uses some of our basic machinery.

Since the unwinding of a structure is a tree, by Theorem 4 Part 1 GCTL$^*$ has the tree model property. By [16, Theorem 6], the graded $\mu$-calculus has the finitely-branching tree model property. Thus it suffices to show that the graded $\mu$-calculus is at least as expressive, over trees, as GCTL$^*$. We do this in two steps.

First, consider counting-CTL$^*$, defined in [19]: it extends CTL$^*$ by path quantifiers of the form $\mathsf{D}^g\phi$ expressing that at least $g$ children of the current node satisfy $\phi$. The graded $\mu$-calculus is at least as expressive as counting-CTL$^*$. This is easily proved by using the fact that the $\mu$-calculus is at least as expressive as CTL$^*$. Indeed, proceed by induction on the structure of counting-CTL$^*$ formulas. Thus the only remaining cases are counting-CTL$^*$ formulas of the form $\mathsf{D}^g\phi$ for $g \in \mathbb{N}$. By induction there is a graded $\mu$-calculus formula $\alpha$ equivalent to $\phi$. Then the graded $\mu$-calculus formula $\langle g \rangle \alpha$ is equivalent to $\mathsf{D}^g\phi$.

Second, by Remark 3, counting-CTL$^*$ is at least as expressive, over trees, as GCTL$^*$. ◀

## B  Appendix to Section 3: Graded Hesitant Tree Automata

### B.1  Alternating Hesitant Tree Automata (AHTA)

For detailed definitions and results regarding automata we refer the reader to: [20] non-deterministic finite word automata (NFW) and non-deterministic Büchi word automata (NBW), [9] for alternating (parity) tree automata, and [17] for alternating hesitant tree automata (AHTA).

We now recall, from the body of the paper, the definition of input tree and run tree for AHTA, and then we make some remarks.

An *input tree (for AHTA)* is a $\Sigma$-labeled $\mathsf{T} = \langle T, V \rangle$ with $T \subseteq D^*$. Note that since $D$ is finite, input trees for AHTA have a fixed finite branching degree. A *run* of an alternating tree automaton $\mathsf{A}$ on an input tree $\mathsf{T} = \langle T, V \rangle$ is a $(T \times Q)$-labeled $\mathbb{N}$-tree $\langle T_r, r \rangle$, such that *(i)* $r(\varepsilon) = (\varepsilon, q_0)$ and *(ii)* for all $y \in T_r$, with $r(y) = (x, q)$, there exists a *minimal* set $\mathsf{S} \subseteq D \times Q$, such that $\mathsf{S} \models \delta(q, V(x))$, and for every $(d, q') \in \mathsf{S}$ there is a son $x \cdot d$ of $x$ and a son $y'$ of $y$ with $r(y') = (x \cdot d, q')$.

Note that if $\delta(q, V(x)) = \mathbf{true}$ then $S = \emptyset$ and the node $y$ has no children; and if there is no $S$ as required (for example if $x$ does not have the required sons) then there is no run-tree with $r(y) = (x, q)$. Observe that disjunctions in the transition relation are resolved into different run-trees, while conjunctions give rise to different sons of a node in a run tree. If $v$ is a node of the run tree, and $r(v) = (u, q)$, call $u$ the *location associated with* $v$, denoted $loc(v)$, and call $q$ the *state associated with* $v$, denoted $state(v)$.

We now discuss the acceptance condition in a little more detail. Fix a run tree $\langle T_r, r \rangle$ and an infinite path $\pi$ in the run tree. Say that the path *visits* a state $q$ at time $i$ if $state(\pi_i) = q$. The structural restriction (i) guarantees that the path $\pi$ eventually gets trapped and visits only states in some element of the partition; i.e., there exists $\mathbb{Q} \in \mathsf{part}$ such that from a certain time $i$ on, $state(\pi_j) \in \mathbb{Q}$ for all $j \geq i$. The condition (ii) ensures that this set is either existential or universal, i.e., $\mathsf{type}(\mathbb{Q}) \in \{exist, univ\}$. Thus, we say that the path $\pi$ *gets trapped in an existential set* if $\mathsf{type}(\mathbb{Q}) = exist$, and otherwise we say that it *gets trapped in a universal set*. We can now define what it means for a path in a run tree to be *accepting*. A path that gets trapped in an existential set is accepting iff it visits some state of $G$ infinitely often, and a path that gets trapped in a universal set is accepting iff it visits every state of $B$ finitely often. Note that the hesitant acceptance condition is a combination of a Büchi and co-Büchi condition and can be thought of as a special case of the parity condition with 3 colors, where the extra structure imposed simply guarantees a space efficient algorithm for deciding membership for these automata [17]. A run $\langle T_r, r \rangle$ of an AHTA is *accepting* iff all its infinite paths are accepting. An automaton $\mathsf{A}$ accepts an input tree $\langle T, V \rangle$ iff there is an accepting run of $\mathsf{A}$ on $\langle T, V \rangle$. The *language* of $\mathsf{A}$, denoted $\mathcal{L}(\mathsf{A})$, is the set of $\Sigma$-labeled $D$-trees accepted by $\mathsf{A}$. We say that an automaton $\mathsf{A}$ is nonempty iff $\mathcal{L}(\mathsf{A}) \neq \emptyset$.

## C Appendix to Section 4: From GCTL* to Graded Hesitant Automata

### Proof of Proposition 8, Page 7

**Proof.** Assume first that $\mathsf{T} \models \mathsf{E}^{\geq g}\psi$, and let $\rho'^1, \ldots, \rho'^g$ be $g$ different paths in the set $\mathsf{min}(Con(\mathsf{S}, \epsilon, \psi))$. For $1 \leq i \leq g$, let $y_i$ be an arbitrarily chosen point on the path $\rho'^i$ satisfying, for every $j \neq i$, that $y_i$ is not on the path $\rho'^j$. Observe that such a point exists since, by minimality, $\rho'^i \not\preceq \rho'^j$ for every $j \neq i$. We thus have that property (i) in the statement of the lemma holds. Property (ii) holds by the minimality of $\rho'^i$. Indeed, the path from the root to the father of $y_i$ is a proper prefix of $\rho'^i$, and is thus not in $Con(\mathsf{S}, \epsilon, \psi)$. By the definition of $\psi$-conservativeness, we have that every path $\rho^i$ in $\mathsf{T}$ such that $\rho'^i \preceq \rho^i$ satisfies $\psi$. Recall that we assume that trees are total, i.e., that they contain no leaves, and thus property (iii) holds by simply taking $\rho^i$ to be any infinite extension of $\rho'^i$.

For the other direction, let $y_1 \ldots, y_g \in T$, be breakpoints satisfying properties (i), (ii), (iii), and consider the paths $\rho^1, \ldots, \rho^g$ through these breakpoints guaranteed by property (iii). For every $1 \leq i \leq g$, let $\rho'^i$ be the shortest prefix of $\rho^i$ such that $\rho'^i$ is $\psi$-conservative, and note that $\rho'^i \in \mathsf{min}(Con(\mathsf{S}, \epsilon, \psi))$. The path $\rho'^i$ is well defined since $\rho^i$ is infinite and satisfies $\psi$ and thus, by definition, it is $\psi$-conservative. In order to prove that $\mathsf{T} \models \mathsf{E}^{\geq g}\psi$, it remains to show that for every $i \neq j$ we have that $\rho'^i \neq \rho'^j$. To see that, observe that for every $1 \leq i \leq g$, property (ii) together with the fact that the property of being $\psi$-conservative is upward closed (with respect to the prefix ordering $\preceq$ of paths), imply that the path from the root to the father of $y_i$ is a proper prefix of $\rho'^i$ and thus, $\rho'^i$ goes through $y_i$. By property (i), if $i \neq j$ then there is no path that goes through both $y_i$ and $y_j$. Combining the last two facts we get that if $i \neq j$ then $\rho'^i \neq \rho'^j$, which completes the proof. ◀

## C.1 Proof of Theorem 9, Page 9

**Proof.** The existence of $\mathbb{A}_\Psi$ is shown in [22, Corollary 23]. The existence of $\mathbb{B}_\Psi$ is proved by a simple adaptation (given below) of the construction in [22, Theorem 22], yielding an alternating finite word automaton $\mathbb{B}'_\Psi$ (of linear size) accepting all finite paths that satisfy $\Psi$, and then converting $\mathbb{B}'_\Psi$ to an equivalent NFW $\mathbb{B}_\Psi$ (using [22, Proposition 16]), of size $2^{O(\Psi)}$.

The alternating finite automaton $\mathbb{B}'_\Psi = \langle \Sigma, Q, q_0, \delta, F \rangle$ is constructed as follows: the input alphabet is $\Sigma = 2^{AP}$ where $AP$ is the set of atoms used by $\Psi$; the set of states $Q$ is the set of all sub-formulas of $\Psi$ and their negations (as usual $\neg\neg\varphi$ is identified with $\varphi$), as well as the special state $eow$ (indicating a guess that we reached the end of the input word); the initial state $q_0$ is $\Psi$; and the set of accepting states $F = \{eow\}$. For a state $\varphi$, and a set of atoms $a$, the transition function $\delta(\varphi, a)$ is given by:

1. if $\varphi = eow$, then $\delta(\varphi, a) = $ **false**;
2. if $\varphi = p$ for $p \in AP$, then $\delta(\varphi, a) = $ **true** if $p \in a$, and $\delta(\varphi, a) = $ **false** otherwise;
3. if $\varphi = \neg p$ for $p \in AP$, then $\delta(\varphi, a) = $ **false** if $p \in a$, and $\delta(\varphi, a) = $ **true** otherwise;
4. If $\varphi = \varphi_1 \dagger \varphi_2$, for $\dagger \in \{\vee, \wedge\}$, then $\delta(\varphi, a) = \delta(\varphi_1, a) \dagger \delta(\varphi_2, a)$;
5. If $\varphi = \neg(\varphi_1 \dagger \varphi_2)$, for $\dagger \in \{\vee, \wedge\}$, then $\delta(\varphi, a) = \delta(\neg\varphi_1, a) \ddagger \delta(\neg\varphi_2, a)$, where $\ddagger$ is the dual of $\dagger$, i.e., $\ddagger = \vee$ if $\dagger = \wedge$, and $\ddagger = \wedge$ if $\dagger = \vee$;
6. if $\varphi = \mathsf{X}\theta$, then $\delta(\varphi, a) = \theta$;
7. if $\varphi = \neg\mathsf{X}\theta$, then $\delta(\varphi, a) = eow \vee \neg\theta$;
8. if $\varphi = \varphi_1 \mathsf{U} \varphi_2$, then $\delta(\varphi, a) = \delta(\varphi_2, a) \vee (\delta(\varphi_1, a) \wedge \delta(\mathsf{X}(\varphi_1 \mathsf{U} \varphi_2), a))$;
9. if $\varphi = \neg(\varphi_1 \mathsf{U} \varphi_2)$, then $\delta(\varphi, a) = \delta(\neg\varphi_2, a) \wedge (\delta(\neg\varphi_1, a) \vee \delta(\neg\mathsf{X}(\varphi_1 \mathsf{U} \varphi_2), a))$.
10. if $\varphi = \varphi_1 \mathsf{R} \varphi_2$, then $\delta(\varphi, a) = (\delta(\varphi_1, a) \wedge \delta(\varphi_2, a)) \vee (\delta(\varphi_2, a) \wedge \delta(\mathsf{X}(\varphi_1 \mathsf{R} \varphi_2), a))$;
11. if $\varphi = \neg(\varphi_1 \mathsf{R} \varphi_2)$, then $\delta(\varphi, a) = \delta(\neg\varphi_2, a) \vee (\delta(\neg\varphi_1, a) \wedge \delta(\neg\mathsf{X}(\varphi_1 \mathsf{R} \varphi_2), a))$.

By defining the transition relation rules for the cases of $\mathsf{X}, \mathsf{U}, \mathsf{R}$ and their negations using one-step unfolding, the adaptation of the construction in [22, Theorem 22] to the finite words semantics addressed here is confined to the definition of the set of accepting states, and the transitions from $eow$ and $\neg\mathsf{X}$. In order to accept $\neg\mathsf{X}\theta$, the automaton can either guess that the input word has ended and go to the accepting state $eow$, or (by going to the state $\neg\theta$) guess that the input has not ended and that the remaining suffix of the input word does not satisfy $\theta$. Having $\delta(eow, a) = $ **false** ensures that if the automaton guessed that the input has ended then any further input would result in a rejecting run. ◀

## C.2 Proof of Lemma 10, Page 8

**Proof.** Given an LTL formula $\Psi$, consider the NBW $\mathbb{A}_\Psi = \langle \Sigma, Q, q_0, \delta, G \rangle$ and the NFW $\mathbb{B}_\Psi = \langle \Sigma, Q', q'_0, \delta', F \rangle$ from Theorem 9. Assume w.l.o.g. that $Q, Q'$ are disjoint (and do not contain $\top, \widetilde{q}_0$) and construct from them a single NBW

$$\mathbb{A}^\Psi = \langle \Sigma, Q \cup Q' \cup \{\top, \widetilde{q}_0\}, \widetilde{q}_0, \delta'', G \cup \{\top\} \rangle,$$

where $\delta''$ is the union of $\delta$ and $\delta'$ as well as the transitions $(\widetilde{q}_0, \sigma, q)$ for every $\sigma$ and $q$ such that $(q_0, \sigma, q) \in \delta$ or $(q'_0, \sigma, q) \in \delta'$; $(\top, \sigma, \top)$ for every letter $\sigma \in \Sigma$, and the transitions $(q, \sigma, \top)$ for every $(q, \sigma, q') \in \delta'$ for which $q' \in F$. I.e., by taking the union of $\mathbb{A}_\Psi$ and $\mathbb{B}_\Psi$, adding a new accepting sink state $\top$, and matching any transition that goes to a final state of $\mathbb{B}_\Psi$ with a transition that goes to the accepting sink $\top$. It is not hard to see that this construction yields the desired automaton. ◀

### C.3 Proof of correctness of the construction of $A_\vartheta$, Page 8

We first give the formal definition of $A_\phi$ in the case that $\phi = \varphi_0 \vee \varphi_1$:

Formally,

$$A_\phi = \langle \Sigma, Q^{\varphi_0} \cup Q^{\varphi_1} \cup \{q_0\}, q_0, \delta, \langle G^{\varphi_0} \cup G^{\varphi_1}, B^{\varphi_0} \cup B^{\varphi_1} \rangle, \langle \mathsf{part}, \mathsf{type}, \preceq \rangle \rangle,$$

where for every $i \in \{0,1\}$, every $\sigma \in \Sigma$, and every $q \in Q^{\varphi_i}$ we have that: $\delta(q, \sigma) = \delta^{\varphi_i}(q, \sigma)$, and $\delta(q_0, \sigma) = \delta^{\varphi_0}(q_0^{\varphi_0}, \sigma) \vee \delta^{\varphi_1}(q_0^{\varphi_1}, \sigma)$; $\mathsf{part} = \{\{q_0\}\} \cup \mathsf{part}^{\phi_1} \cup \mathsf{part}^{\phi_2}$; $\mathsf{type}(\{q_0\}) = trans$, and for $i \in \{1,2\}$, and every $\mathbb{Q} \in \mathsf{part}^{\phi_i}$, we have $type(\mathbb{Q}) = type^{\phi_i}(\mathbb{Q})$; and $\preceq$ is the union of the relations $\preceq^{\phi_1}, \preceq^{\phi_2}$ as well as the inequalities $\mathbb{Q} \preceq \{q_0\}$ for every $\mathbb{Q} \in \mathsf{part}$. In words, we maintain the partitioning of the states of $Q^{\varphi_0}$ and $Q^{\varphi_1}$ their types and order, and add the transient set $\{q_0\}$ making it larger then all other sets.

Before proving the correctness of the (entire) construction, we need some notation and a lemma. Let $A_\phi$ be the automaton constructed for a formula of the form $\phi = E^{\geq g}\psi$, and let $\langle T_r, r \rangle$ be a run of $A_\phi$ on an input tree $T = \langle T, V \rangle$. Given a node $v \in T_r$, and its label $r(v) = (u, q)$, we say that $i$ is active (disabled) in $v$ iff $state(v) \in Q_1$ and $state(v)_i \neq \bot$ ($state(v)_i = \bot$).

▶ **Lemma 17.** *Let $A_\phi$ be the automaton constructed for $\phi = E^{\geq g}\psi$, and let $\langle T_r, r \rangle$ be a run of $A_\phi$ on a tree $T = \langle T, V \rangle$. For every $1 \leq i \leq 2g$, the set of nodes of $T_r$ in which $i$ is active forms an infinite path $\pi^i$ from the root. Furthermore, if $\langle T_r, r \rangle$ is an accepting run, then for $1 \leq i \leq g$, there is an $i_k > 1$ such that $state(\pi^i_{i_k-1})_{i+g} \in Q^\neg \setminus \{\top\}$, and for every $l \geq i_k$ the only active $\Psi$ coordinate in $state(\pi^i_l)$ is $i$.*

**Proof of Lemma 17.** We first prove that the set of nodes $I$ of $T_r$ in which $i$ is active is an infinite path from the root. The proof is by induction on the depth $\kappa$ of the nodes in $I$. The induction hypothesis is that there is exactly one node of depth $\kappa$ in $I$, and that for $\kappa \geq 1$ it is a son of a node in $I$. For the base case $\kappa = 0$, the root $\varepsilon$ of $T_r$ satisfies $state(\varepsilon) = q_0$, and note that all coordinates, and in particular the $i$'th coordinate, are active in the initial state $q_0$.

For $\kappa > 1$, assume that the induction hypothesis holds. First, note that, by the definition of $\delta$ (and in particular property (i) of a legal distribution), it must be that there is at least one node of depth $\kappa$ in $I$. Assume by way of contradiction that there are two such nodes $y \neq y' \in I$ of depth $\kappa$. Observe that the transition function $\delta$ is such that once a coordinate is disabled it can never become active again. Hence, the parents $x$ of $y$ and $x'$ of $y'$ both have the $i$'th coordinate active. Thus, by the induction hypothesis, $x = x'$, and $y$ and $y'$ are siblings. Let $r(x) = (s^x, q^x), r(y) = (s^y, q^y), r(y') = (s^{y'}, q^{y'})$, and let $d$ be the number of children of $s^x$. Note that the definition of a run tree implies that the formula $expand_d(\delta(q^x, V(s^x)))$ contains a conjunction having both $(d, q^y)$ and $(d', q^{y'})$, where $d, d'$ are the directions in the input tree assigned to $s^y, s^{y'}$ respectively. In other words, the copy of $A_\phi$ in state $q^x$, that reads the input node $s^x$, launches (at least) two copies *in parallel*: one in state $q^y$ to the son $s^y$, and one in state $q^{y'}$ to the son $s^{y'}$. Observe that all transitions from states in $Q_1$, and thus in particular from $q^x$, are of the form $\vee_{\sigma' \in \Sigma'}((\vee_{X \in Legal(q,\sigma')} \Diamond(X)) \wedge \Omega_{\sigma'})$, where $\Omega_{\sigma'}$ is a boolean formula that involves only states in $Q_2$. But this is a contradiction since this implies that for some $\Diamond(X)$ in this transition we have that $q^y, q^{y'} \in X$, which is impossible by property (ii) of a legal distribution.

We now prove that if $T_r$ is accepting, then for $1 \leq i \leq g$ an index $i_k$ as stated by the lemma exists. Since $\pi^i$ is an infinite path in $T_r$, and all states in $Q_1$ belong to existential sets (i.e. sets with type = $exist$), the fact that $T_r$ is an accepting run implies that for infinitely many $l's$ we have that $state(\pi_l^i) \in G$. Note that all states in $G$ have only one active $\Psi$ coordinate, and that once a coordinate becomes disabled it is never enabled again. Thus, there is a minimal index $i_k$ such that for every $l \geq i_k$ the only active $\Psi$ coordinate in $state(\pi_l^i)$ is $i$. The fact that $state(\pi_{i_k-1}^i)_{i+g} \in Q^\neg \setminus \{\top\}$ follows immediately from the minimality of $i_k$ and property (iii) in the definition of a legal distribution. ◂

**Proof of Theorem 7.** The proof is by induction on the structure of $\vartheta$ and shows that, at each stage of the construction, for every sub-formula $\phi$ of $\vartheta$, the automaton $\mathsf{A}_\phi$ satisfies the statement of the theorem. The depth, number of states, and size of the transition function are already computed in Proposition 11.

We begin by showing that if $\langle T_r, r \rangle$ is an accepting run of $\mathsf{A}_\phi$ on a tree $\mathsf{T} = \langle T, V \rangle$, then $\mathsf{T} \models \varphi$. The cases $\phi = p$, $\phi = \varphi_0 \lor \varphi_1$, and $\phi = \neg \varphi$ follow easily from the definitions and the induction hypothesis. Consider now the case $\phi = \mathsf{E}^{\geq g} \psi$, and for every $1 \leq i \leq g$, let $\pi^i$, and $i_k$ be as given by Lemma 17; furthermore, let $r(\pi_{i_k}^i) = (y_i, q^i)$, and take $y_1, \ldots, y_g$ to be the breakpoints in the statement of Proposition 8. We claim that the conditions of Proposition 8 are satisfied, and thus, it's conclusion also holds, i.e., that $\mathsf{T} \models \varphi$ as required.

First, consider condition (i) of Proposition 8: given $i \neq j$, we have to show that $y_i$ is not a descendant of $y_j$. Observe that since $i$ is the only active $\Psi$ coordinate in $\pi_{i_k}^i$, and $j$ is the only active $\Psi$ coordinate in $\pi_{j_k}^j$, then $\pi_{i_k}^i \neq \pi_{j_k}^j$. Let $x \in \mathsf{T}_r$ be a node of maximal depth $\kappa$, in which both coordinates $i$ and $j$ are active (note that $x$ is well defined since both coordinates are active at the root), and let $r(x) = (s^x, q^x)$. By Lemma 17, $x$ must be a common ancestor of both $\pi_{i_k}^i$ and $\pi_{j_k}^j$, and $\pi_{\kappa+1}^i, \pi_{\kappa+1}^j$ are thus sons of $x$. Let $r(\pi_{\kappa+1}^i) = (s', q'), r(\pi_{\kappa+1}^j) = (s'', q'')$, and note that by the maximality of $\kappa$, in $q'$ coordinate $i$ is active and $j$ is not, and vice versa for $q''$. Hence, $q' \neq q''$. In other words, the copy of $\mathsf{A}_\phi$ in state $q^x$, that reads the input node $s^x$, launches (at least) two *different* copies in parallel: one in state $q'$ to $s'$, and the other in state $q''$ to $s''$. Recall that the transition $\delta(q^x, V(s^x))$ is of the form $\lor_{\sigma' \in \Sigma'}((\lor_{X \in Legal(q, \sigma')} \Diamond(X)) \land \Omega_{\sigma'})$, where $\Omega_{\sigma'}$ is a boolean formula that involves only states in $Q_2$ (and thus not $q'$ and $q''$). By the definition of a run, $\langle T_r, r \rangle$ makes use of a single disjunct of any disjunction, and thus in this case, of one $\Diamond(X)$. It follows that both $q'$ and $q''$ appear in $X$, and thus, by the semantics of $\Diamond$, it must be that $q'$ and $q''$ were sent to two different sons of $s^x$, i.e., that $s' \neq s''$. Recall that $i_k \geq \kappa + 1$, and $j_k \geq \kappa + 1$, and thus $y_i$ is either equal to $s'$ or is a descendant of it. Similarly, $y_j$ is either equal to $s''$ or is a descendant of it. We conclude that $y_i$ is not a descendant of $y_j$ as needed.

We now address condition (ii) of Proposition 8. Given $1 \leq i \leq g$, let $m = i + g$, and take the path $\pi^m$ guaranteed by Lemma 17. Consider the path $\rho^m = loc(\pi_0^m) \cdot loc(\pi_1^m) \cdots$ in $T$, of the nodes associated with $\pi^m$. For every $l \geq 0$, let $\sigma_l' \in \Sigma'$ be the set of maximal state subformulas of $\phi$ that hold in $\rho_l^m$. Applying the induction hypothesis to all $\theta \in \overline{max(\phi)}$, we can conclude that the only way $T_r$ can be accepting is if for every $0 \leq l$ it resolves the outermost disjunction in $\delta(state(\pi_l^m), V(\rho_l^m))$ by taking the disjunct

$$(\lor_{X \in Legal(state(\pi_l^m), \sigma_l')} \Diamond(X)) \land (\land_{\theta \in \sigma_l'} \delta^\theta(q_0^\theta, \sigma)) \land (\land_{\theta \notin \sigma_l'} \delta^{\neg \theta}(q_0^{\neg \theta}, \sigma)).$$

It is thus not hard to see that since $T_r$ is an accepting run of $\mathsf{A}_\phi$ on $\mathsf{T}$, then

$$r' = state(\pi_0^m)_m, state(\pi_1^m)_m, \ldots$$

is an accepting run of $\mathbb{A}^{\neg \Psi}$ on the word $w = \sigma_0' \cdot \sigma_1' \cdots \in \Sigma'^\omega$. Note that Lemma 17 implies that $\pi_{i_k-1}^i = \pi_{i_k-1}^m$, and it also states that $state(\pi_{i_k-1}^i)_{i+g} \neq \top$. Hence, by Theorem 10, some (finite or infinite) prefix $u = \sigma_0' \cdot \sigma_1' \cdots$ of $w$ of length at least $i_k$ satisfies $\neg \Psi$. By Lemma 3, it follows that the prefix $\varrho$ of $\rho^m$, of the same length as $u$, satisfies $\neg\psi$. Observe that the length of $\varrho$ implies that the path $\rho_0^m \cdots \rho_{i_k-1}^m$, from the root of $T$ to the father of $y_i$, is a prefix (possibly not a proper prefix) of $\varrho$, and is thus not $\psi$-conservative, as required by condition (ii).

Addressing condition (iii) of Proposition 8 follows in the footsteps of the reasoning used for condition (ii). Given $1 \leq i \leq g$ and the path $\pi^i$, the associated path $\rho^i = loc(\pi_0^i) \cdot (\pi_1^i) \cdots$ in $T$ induces the infinite word $w'' = \sigma_0'' \cdot \sigma_1'' \cdots$ who's letters are the sets of maximal state subformulas of $\psi$ that hold along the path $\rho^i$. By the induction hypothesis, and since $T_r$ is accepting, the run $state(\pi_0^i)_i, state(\pi_1^i)_i, \ldots$ is an accepting run of $\mathbb{A}_\Psi$ on the word $w''$, and thus by Lemma 3, the path $\rho^i$ satisfies $\psi$. Since $y_i$ lies on $\rho^i$, condition (iii) of Proposition 8 is met, and we can conclude that $\mathsf{T} \models \varphi$.

For the other direction, let $\mathsf{T} = \langle T, V \rangle$ be such that $\mathsf{T} \models \varphi$. We have to show that $\mathsf{A}_\phi$ has an accepting run $\langle T_r, r \rangle$ on $\mathsf{T}$. As before, the cases $\phi = p$, $\phi = \varphi_0 \vee \varphi_1$, and $\phi = \neg\varphi$ follow easily from the definitions and the induction hypothesis. For the case $\phi = \mathsf{E}^{\geq g}\psi$, by Proposition 8, there are $g$ breakpoints $y_1, \ldots, y_g \in T$, and rooted infinite paths $\rho^1, \ldots, \rho^g$, such that for every $1 \leq i \leq g$ we have that $y_i = \rho_{i_k}^i$ for some $i_k \geq 1$, and $\rho^i \models \psi$; furthermore, the prefix $\rho_0^i, \ldots, \rho_{i_k-1}^i$ is not $\psi$-conservative, and thus, it can be extended to an infinite path $\rho^{i+g}$ such that some (finite or infinite) prefix of $\rho^{i+g}$ of length $i_n \geq i_k$ satisfies $\neg\psi$. For every node $x \in T$, let $\sigma'(x) \subseteq max(\phi)$ be the set of all maximal state subformulas of $\psi$ that hold in $x$. By Lemma 3, for every $1 \leq i \leq g$, the infinite word $w^i = \sigma'(\rho_0^i) \cdot \sigma'(\rho_1^i) \cdots$ satisfies $\Psi$, and the (finite or infinite) word $w^{i+g} = \sigma'(\rho_0^{i+g}) \cdots \sigma'(\rho_{i_n-1}^{i+g})$ satisfies $\neg\Psi$. By Theorem 9 there is an accepting run $r^i$ of $\mathbb{A}_\Psi$ on $w^i$; and by Lemma 10 there is an accepting run $r^{i+g}$ of $\mathbb{A}^{\neg\Psi}$ on $w^{i+g}$ for which $r_j^{i+g} \neq \top$ for all $j < i_n$.

We build an accepting run $\langle T_r, r \rangle$ of $\mathsf{A}_\phi$ on $\mathsf{T}$, by induction on the depth $\kappa$ of the node $x \in T$. At the root $\varepsilon$ of $T$, the automaton is in the initial state $q_0$, and note that $q_0 \in Q_1$. For the induction step, a copy of $\mathsf{A}_\phi$ in some state $q \in Q_1$, that is at a node $x \in T$ of depth $\kappa \geq 0$ whose labelling $V(x) = \sigma$, proceeds as follows. Recall that

$$\delta(q, \sigma) = \vee_{\sigma' \in \Sigma'} \left[ (\vee_{X \in Legal(q,\sigma')} \Diamond(X)) \wedge (\wedge_{\theta \in \sigma'} \delta^\theta(q_0^\theta, \sigma)) \wedge (\wedge_{\theta \notin \sigma'} \delta^{\neg\theta}(q_0^{\neg\theta}, \sigma)) \right].$$

First, the automaton resolves $\vee_{\sigma' \in \Sigma'}$ by choosing $\sigma' = \sigma'(x)$. By the induction hypothesis (of the theorem), we know that for every $\theta \in \sigma'$ (alternatively $\theta \notin \sigma'$) the automaton $\mathsf{A}_\theta$ (alternatively $\mathsf{A}_{\neg\theta}$) has an accepting run on the subtree of $\mathsf{T}$ rooted at $x$; thus, by following these accepting runs, $\mathsf{A}_\phi$ can satisfy the conjunction $(\wedge_{\theta \in \sigma'} \delta^\theta(q_0^\theta, \sigma)) \wedge (\wedge_{\theta \notin \sigma'} \delta^{\neg\theta}(q_0^{\neg\theta}, \sigma))$. It remains to show how the automaton handles $\vee_{X \in Legal(q,\sigma')} \Diamond(X)$. For every node $t \in T$, let $live(t) = \{1 \leq i \leq g \mid t \in \rho^i\}$ be the set of all $i$'s for which the path $\rho^i$ goes through $t$. Let $s_1, \ldots, s_m$ be the sons of $x$ for which these $live()$ sets are not empty. Let $Y = q^1, \ldots, q^m$ be a sequence of states where for every $1 \leq h \leq m$, and every $1 \leq i \leq 2g$, we have that $q_i^h = r_{k+1}^i$ if $i \in live(s_h)$, and $q_i^h = \bot$ otherwise. In words, the $i$'th coordinate of $q^h$ follows the run $r^i$ if $\rho^i$ goes through the $h$'th son of $x$, and is disabled if $\rho^i$ does not go through this son. We claim that $Y$ is a legal distribution of $(q, \sigma')$. Indeed, it is not hard to see that $q^1, \ldots, q^m$ are all different, and that properties (i), (ii) and (iv) of a legal distribution are satisfied. As for property (iii), recall that by Proposition 8, if $j, l \in \{1, \ldots, g\}$ and $j \neq l$, then $y_j$ is not a descendant of $y_l$ (and vice-versa). Thus, if both $j$ and $l$ are active in $q^h$ (i.e., both $\rho^j$ and $\rho^l$ go through $s_h$), it must be that $y_j$ and $y_l$ are both descendants of $s_h$, and thus $k+1 < j_k$ and $k+1 < l_k$. Recall that for every $1 \leq i \leq g$, the paths $\rho^i$ and $\rho^{i+g}$ coincide at least up to (and including) the father of $y_i$. Hence, coordinates $j+g$, and $l+g$ must also be active in $q^h$. Also, recall that for every $1 \leq i \leq g$ we have that $r_m^{i+g} \neq \top$ for all $m < i_n$, and that $i_n \geq i_k$. Thus, in particular, $r_{k+1}^{j+g} = q_{j+g}^h \neq \top$ and $r_{k+1}^{l+g} = q_{l+g}^h \neq \top$, and property (iii) holds, and $Y$ is a legal distribution of $(q, \sigma')$. Hence, the automaton can handle $\vee_{X \in Legal(q,\sigma')} \Diamond(X)$ by taking $\Diamond(Y)$, and resolving $\Diamond(Y)$ by sending, for every $1 \leq h \leq m$, a copy in state $q^h$ to the son $s_h$ of $x$.

We now argue that the run $\langle T_r, r \rangle$ described above is accepting. Let $\pi$ be a path in the run tree, and consider the case that for some $j$ we have that $state(\pi_j) \in Q_2$. Take $j$ to be minimal with this property, and observe that it must be that $state(\pi_j) \in Q^\theta$ for some $\theta \in \overline{max(\phi)}$. By our construction of $\langle T_r, r \rangle$, the subtree rooted at $\pi_j$ is an accepting run of $A_\theta$ on the subtree of $T$ rooted at $loc(\pi_j)$, and thus $\pi$ is an accepting path. Consider now paths for which all states associated with the nodes of the path are in $Q_1$. By Lemma 17, there are exactly $2g$ such paths $\pi^1, \ldots, \pi^{2g}$, and it is easy to see that by our construction of the run, for every $1 \leq i \leq 2g$, we have that $state(\pi_1^i)_i, state(\pi_2^i)_i, \ldots$ is exactly the run $r^i$, and that for every $j \geq i_n$ the only active $\Psi$ coordinate in $state(\pi_j^i)$ is $i$. Hence, by the definition of the acceptance condition of $A_\phi$, the path $\pi^i$ is accepting. Which completes the correctness proof of the construction. ◀

## C.4 Proof of Proposition 11, Pages 10

**Proof.** The depth is clearly $O(|\vartheta|)$.

We analyse the number of states, by cases. As usual, the cases $\vartheta = p$, $\vartheta = \varphi_0 \vee \varphi_1$, and $\vartheta = \neg\varphi$ follow easily from the definitions and the induction hypothesis. For the case $\vartheta = E^{\geq g}\psi$, the states of the automaton are the union of $Q_1$ and $Q_2$.

The set $Q_2$ uses states from each automaton $A_\theta$ for every $\theta \in \overline{max(\phi)}$, and thus $|Q_2|$ is $\Sigma_{\theta \in \overline{max(\psi)}} |A_\theta|$. But by induction $|A_\theta|$ is $2^{O(|\theta| \times \deg(\theta))}$, and so $|Q_2|$ is at most $O(|\psi|) \times 2^{O(|\psi| \times \deg(\psi))} = 2^{O(|\psi| \times \deg(\psi))}$.

The set $Q_1$ uses a vector of $g$ copies of $\mathbb{A}_\Psi$ and $g$ copies of $\mathbb{A}^{\neg\Psi}$. Thus $|Q_1|$ is $|\mathbb{A}_\Psi|^g \times |\mathbb{A}^{\neg\Psi}|^g$. Since $|\mathbb{A}_\Psi|$ and $|\mathbb{A}^{\neg\Psi}|$ are $2^{O(|\psi|)}$, we get that $|Q_1|$ is $2^{O(|\psi| \times g)} = 2^{O(|\vartheta| \times \deg(\vartheta))}$

Thus, the number of states of $A_\vartheta$ is $|Q_1| + |Q_2|$ which is $2^{O(|\vartheta| \times \deg(\vartheta))}$.

We treat the size of the transition function similarly. For the case $\phi = E^{\geq g}\psi$ we add the lengths of the transitions leaving $Q_1$, and the lengths of the transitions leaving $Q_2$.

Say $q \in Q_2$ and $\sigma \in \Sigma$, and let $\theta \in \overline{max(\psi)}$ be such that $q \in Q^\theta$. Then the length of formula $\delta(q, \sigma)$ is, by induction, at most $2^{O(|\theta| \times \deg(\theta))}$. Thus the length of the transitions leaving $Q_2$ is at most $|Q_2| \times 2^{O(|\psi| \times \deg(\psi))} = 2^{O(|\vartheta| \times \deg(\vartheta))}$.

Say $q \in Q_1$ and $\sigma \in \Sigma$. By induction, for $\theta \in \overline{max(\psi)}$, the number of transitions in $A_\theta$ is $2^{O(|\theta| \times \deg(\theta))}$. Then the transition $\delta(q, \sigma)$, defined as,

$$\bigvee_{\sigma' \in 2^{max(\psi)}} \left[ (\bigvee_{X \in Legal(q,\sigma')} \Diamond(X)) \bigwedge (\wedge_{\theta \in \sigma'} \delta^\theta(q_0^\theta, \sigma)) \bigwedge (\wedge_{\theta \notin \sigma'} \delta^{\neg\theta}(q_0^{\neg\theta}, \sigma)) \right]$$

has length at most

$$2^{O(|\psi|)} \times \left[ (\sum_{X \in Legal(q,\sigma')} |\Diamond(X)|) + (\Sigma_{\theta \in \overline{max(\psi)}} 2^{O(|\vartheta| \times \deg(\vartheta))}) \right].$$

Now $Legal(q, \sigma')$ is the number of legal distributions of $q$, which is at most the number of ways each of the $2g$ states can evolve times the number of ways to partition the components of $q$ into $k$ pieces (for some $k \leq 2g$). This is at most $(2^{O(|\psi|)})^{2g} \times 2g \times (2g)^{2g}$ which is at most $2^{O(|\psi| \times \deg(|\psi|))}$.

Also, by writing $X$ as a $2g$-tuple of states of $Q_1$ in which each co-ordinate is also given a number (between 1 and $k$) indicating which element of the partition it is in, we get that $|\Diamond(X)|$ is at most $|Q_1|^{2g} + k2g \leq |Q_1|^{2g} + (2g)^2 = 2^{O(|\vartheta| \times \deg(|\vartheta|))} + O(\deg(|\vartheta|)^2) = 2^{O(|\vartheta| \times \deg(|\vartheta|))}$.

Thus the length is at most

$$2^{O(|\psi|)} \times \left[ 2^{O(|\psi| \times \deg(|\psi|))} \times 2^{O(|\vartheta| \times \deg(|\vartheta|))} + O(|\psi|) \times 2^{O(|\vartheta| \times \deg(\vartheta))} \right]$$

which is $2^{O(|\vartheta| \times \deg(\vartheta))}$. ◀

## D Appendix to Section 5: Complexity of Satisfiability and Model-checking of GCTL*

### Proof of Theorem 12, Page 10

**Proof.** Suppose $\vartheta$ is satisfiable. By Theorem 4, $\vartheta$ has a finitely-branching tree model. We will prove that every tree model has a subtree of branching degree $|Q|^2$ that is also a model of $\vartheta$, where $Q$ is the state set of the automaton $\mathsf{A}_\vartheta$. Since by Theorem 7 we have that $|Q| = 2^{O(|\vartheta| \times \deg(\vartheta))}$, the theorem follows.

Our proof makes use of the automaton $\mathsf{A}_\vartheta$. Recall that is built recursively on the state subformulas and their negations $\phi$ of $\vartheta$. In stage $\phi$ an automaton is built which consists of some new states as well as the states of automata built from subformulas of $\phi$ and their negations. Let $\phi_q$ denote the stage at which state $q$ enters the construction for the first time. Note that every state of $Q$ enters the construction at some time, but some created states are not part of $Q$ (for example, the no state of the automaton for $(\neg p) \lor q$ is created at the stage corresponding to $p$).

We use the following standard notion. The membership game $G_{\mathsf{T},\mathsf{A}_\vartheta}$ of tree $\mathsf{T}$ and the automaton $\mathsf{A}_\vartheta$ is played by two players, *automaton* and *pathfinder*. Player automaton moves by resolving disjunctions, and is trying to show that $\mathsf{T}$ is accepted by $\mathsf{A}_\vartheta$. Player pathfinder moves by resolving conjunctions and is trying to show that $\mathsf{T}$ is not accepted by $\mathsf{A}_\vartheta$. As usual, player automaton has a winning strategy if and only if $\mathsf{T} \models \mathsf{A}_\vartheta$. By memoryless determinacy of parity games on infinite arenas, player automaton has a winning strategy if and only if he has a memoryless winning strategy. This memoryless property will be used to construct the required subtree.

**We now define $G_{\mathsf{T},\mathsf{A}_\vartheta}$ (for tree $\mathsf{T} = \langle T, V \rangle$ and formula $\vartheta$).** The arena consists of the *main nodes* $Q \times T$, two *sink nodes* $\top, \bot$, as well as *auxilliary nodes* which are used to play the auxilliary games $aux(q,t)$ for $(q,t) \in Q \times T$. Play proceeds from a main node $(q,t)$ to the auxilliary arena $aux(q,t)$ (formally defined below) played on the parse tree of the formula defined by $\delta(q, V(t))$. The auxilliary arena $aux(q,t)$ is a finite tree, and when a play $\pi$ exits this arena it results in a node $exit_\pi(q,t)$ which is either a main node from $(Q \times sons(t)) \cup (Q \times \{t\})$ or a sink node. A play $\pi$ that visits $(q,t)$, proceeds, via some auxilliary nodes and main nodes of the form $Q \times \{t\}$, to a node $next_\pi(q,t) \in (Q \times sons(t)) \cup \{\bot, \top\}$.

The definition of the game $aux(q,t)$ depends on the form of $\phi_q$ and the definition of the transition $\delta(q, V(t))$.

- If $\phi_q = p$ for $p \in \mathrm{AP}$, then the game $aux(q,t)$ immediately results in sink node $\top$ if $p \in V(t)$ and in sink node $\bot$ otherwise.
- If $\phi_q = \varphi_0 \lor \varphi_1$, then in the game $aux(q,t)$ automaton chooses to exit either to main node $(q_0^{\varphi_0}, t)$ or to main node $(q_0^{\varphi_1}, t)$.
- If $\phi_q = \neg \varphi$, then the game $aux(q,t)$ immediately results in main node $(q', t)$ where $q'$ is the initial state of the dual automaton for $\mathsf{A}_\varphi$.

- If $\phi_q = \mathsf{E}^{\geq g}\psi$, then the game $aux(q,t)$ proceeds as follows: first player automaton picks $\sigma' \in \Sigma'$, and then pathfinder has three choices. Either she i) picks $\theta \in \sigma'$ and exits at main node $(q_0^\theta, t)$, or ii) she picks $\theta \notin \sigma'$ and exits at main node $(q_0^{\neg\theta}, t)$, or iii) she transfers play to automaton, in which case automaton picks a legal distribution, say $X = (q_1, \cdots, q_m) \in Legal(q, \sigma')$, and automaton also picks $m$-many different sons of $t$, say $(s_1, \cdots, s_m)$, and then pathfinder picks some $i \leq m$, and exits at main node $(q_i, s_i)$. To understand this game, recall from the construction of $\mathsf{A}_\phi$ that the transition relation for this case is defined as

$$\bigvee_{\sigma' \in 2^{max(\psi)}} \left[ (\bigvee_{X \in Legal(q,\sigma')} \Diamond(X)) \bigwedge (\wedge_{\theta \in \sigma'} \delta^\theta(q_0^\theta, \sigma)) \bigwedge (\wedge_{\theta \notin \sigma'} \delta^{\neg\theta}(q_0^{\neg\theta}, \sigma)) \right].$$

The hesitant acceptance condition of $\mathsf{A}_\vartheta$ can be easily translated into a parity condition with priorities $\{0, 1, 2\}$ (also, let sink node $\top$ have priority 2, and sink node $\bot$ have priority 1). We say that player automaton wins a play if the largest priority occuring infinitely often is even.

This completes the description of the membership game $G_{\mathsf{T},\mathsf{A}_\vartheta}$.

We now continue with the proof. Since $\vartheta$ is satisfiable, it is satisifiable by some finitely-branching tree $\mathsf{T}$. Thus, fix a memoryless winning strategy $str$ for player automaton in the game $G_{\mathsf{T},\mathsf{A}_\vartheta}$.

**Lemma** (†). For every main node $(q,t)$ of $G_{\mathsf{T},\mathsf{A}_\vartheta}$, there exists a set $Y(q,t) \subseteq Q \times sons(t)$ such that

- $|Y(q,t)| \leq |Q|$, and
- every play $\pi$ consistent with $str$ that exits the arena $aux(q,t)$ with $exit_\pi(q,t) \in Q \times sons(t)$ actually satisfies that $exit_\pi(q,t) \in Y(q,t)$.

Assume for now that Lemma (†) is true. Then every play consistent with $str$ only visits, besides the main nodes $Q \times \{root\}$ (here $root$ is the root vertex of tree $\mathsf{T}$), main nodes from $\cup_{t \in T} X(t)$ where $X(t) := \cup_{q \in Q} Y(q,t)$ (for $t \in T$). Note that for all $t \in T$, $|X(t)| \leq |Q|^2$. Define the subtree $\mathsf{T}'$ of $\mathsf{T}$ where the domain $T'$ consists of $root$ and the elements in the set $\{t \in T : \exists q \in Q.(t,q) \in X(t)\}$. Note that every node in $\mathsf{T}'$ has degree at most $|Q|^2$. The membership game $G_{\mathsf{T}',\mathsf{A}_\vartheta}$ is a subgame of $G_{\mathsf{T},\mathsf{A}_\vartheta}$, and player automaton's strategy $str$ is well defined on this subgame, and is winning. Thus $\mathsf{T}' \models \vartheta$.

**Finally, we prove Lemma** (†) **by induction on** $\phi$: for every state $q$ such that $\phi_q = \phi$, for every $t \in T$, there exists a set $Y^\phi(q,t) \subseteq Q^\phi \times sons(t)$ of size at most $|Q^\phi|$ such that every play $\pi$ consistent with $str$ that exits the auxilliary arena $aux(q,t)$ in a node of the form $Q^\phi \times sons(t)$ actually exits it in a node from $Y^\phi(q,t)$.

To see why this gives the lemma, take $(q,t) \in Q \times T$, and consider the induction at stage $\phi_q$. Then every play $\pi$ that exits the arena $aux(q,t)$ with $exit_\pi(q,t) \in Q \times sons(t)$ actually satisfies that $exit_\pi(q,t) \in Y^{\phi_q}(q,t)$. But $Y^{\phi_q}(q,t) \subseteq Q \times sons(t)$ and $|Y^{\phi_q}(q,t)| \leq |Q^{\phi_q}| \leq |Q|$.

For the proof, suppose every proper subformula of $\phi$ satisfies the inductive hypothesis. There are four cases:

- $\phi = p$ for some $p \in \mathrm{AP}$. Define $Y^\phi(q,t) := \emptyset$, and note that the exit node of $aux(q,t)$ is a sink node.
- $\phi = \varphi_1 \vee \varphi_2$. Define $Y^\phi(q,t) := \emptyset$, and note that the exit node of $aux(q,t)$ is a main node of the form $Q \times \{t\}$.
- $\phi = \neg\varphi$. Define $Y^\phi(q,t) := \emptyset$, and note that the exit node of $aux(q,t)$ is a main node of the form $Q \times \{t\}$.

- $\phi = \mathsf{E}^{\geq g}\psi$. The only way to exit $aux(q,t)$ in a main node of the form $Q \times sons(t)$ is via option iii) in the definition of $aux(q,t)$ above; i.e., automaton picks $\sigma'$, and then pathfinder transfers play to automaton, who then, according to $str$, picks a legal distribution, say $X = (q_1, \cdots, q_m)$, and corresponding sons of $t$, say $(s_1, \cdots, s_m)$, and then pathfinder picks an exit of the form $(q_i, s_i)$. Define $Y^\phi(q,t) := \{(q_1, s_1), \cdots, (q_m, s_m)\}$. Since $m \leq |Q^\phi|$ (the components of the legal distribution $X$ are distinct elements of $Q^\phi$), we have that $|Y^\phi(q,t)| \leq |Q^\phi|$.

This completes the proof of the lemma, and hence of the theorem. ◀

## D.1 Proof of Theorem 14, Page 11

**Proof.** The lower bound follows from the corresponding lower-bound for CTL$^*$ [23]. For the upper bound, given a GCTL$^*$ formula $\vartheta$, using Theorem 7 construct the GHTA $\mathsf{A}_{\neg\vartheta}$, which has $2^{O(|\vartheta| \times \mathsf{deg}(\vartheta))}$ states, and transition function of size $2^{O(|\vartheta| \times \mathsf{deg}(\vartheta))}$, and which has depth $O(|\vartheta|)$. Let $d$ be the largest degree in $\mathsf{S}$.

Claim: For every GHTA $\mathsf{A} = \langle \Sigma, D, Q, q_0, \delta, \langle G, B \rangle, \langle \mathsf{part}, \mathsf{type}, \preceq \rangle \rangle$ and $d \in \mathbb{N}$, there is an AHTA $\mathsf{A}' = \langle \Sigma, Q, q_0, \delta', \langle G, B \rangle, \langle \mathsf{part}, \mathsf{type}, \preceq \rangle \rangle$ that accepts the same $d$-ary trees as $\mathsf{A}$, and such that $||\delta'|| \leq ||\delta|| \times |Q|^d$.

Proof of Claim: When only considering trees with a bounded branching degree $d$, one can transform a GHTA $\mathsf{A}$ into an equivalent AHTA $\mathsf{A}'$ by simply replacing $\delta(q,\sigma)$ by $expand_d(\delta(q,\sigma))$. Note that expanding the $\diamond_Q$s and $\square_Q$s only blows up the size of the transition relation by a multiplicative factor of $|Q|^d$, and leaves the state space unchanged. This completes the proof of the Claim.

Apply the Claim to get an equivalent AHTA $\mathsf{A}'$ of size $||\mathsf{A}'|| = d + (2^{O(|\vartheta| \times \mathsf{deg}(\vartheta))} \times |Q|^d) + 2^{O(|\vartheta| \times \mathsf{deg}(\vartheta))} = 2^{O(|\vartheta| \times \mathsf{deg}(\vartheta) + d \times |\vartheta| \times \mathsf{deg}(\vartheta))}$, and of depth $\partial = O(|\vartheta|)$. By Theorem 5, the membership problem of the AHTA $\mathsf{A}'$ on LTS $\mathsf{S}$ can be solved in space $O(\partial \log^2(|S| \times ||\mathsf{A}'||))$ which is polynomial in $|\vartheta|$ and $|S|$ (using $\mathsf{deg}(\vartheta) \leq |\vartheta|$ and $d \leq |S|$). ◀

# E Appendix to Discussion

▶ **Theorem 18.** *The satisfiability problem for* GCTL *is* ExpTime-Complete. *The model checking problem for* GCTL *and finite LTS is in* PTime.

**Proof.** Consider the construction in Theorem 7 of $\mathsf{A}_\vartheta$ when $\vartheta$ it taken from the fragment GCTL of GCTL$^*$, and in particular where it comes to a subformula $\phi$ of the form $\phi = \mathsf{E}^{\geq g}\psi$. Since $\psi$ is either of the form $p\mathsf{U}q$ or $\mathsf{X}p$, the number of new states added at this stage is a constant. Thus, the number of states of $\mathsf{A}_\vartheta$ is linear in the size of $\vartheta$. ◀